\documentclass[prl,preprintnumbers,amsmath,amssymb]{revtex4}

\usepackage{mathrsfs}
\usepackage{graphicx}
\usepackage{dcolumn}
\usepackage{bm}

\begin{document}

\title{Maxwell's demon in biochemical signal transduction with feedback loop}

\author{Sosuke Ito$^1$ and Takahiro Sagawa$^{2}$}
 
\affiliation{$^1$Department of Physics, The University of Tokyo, Tokyo 113-0033, Japan (email: sosuke@daisy.phys.s.u-tokyo.ac.jp)\\
$^2$Department of Basic Science, The University of Tokyo, Tokyo 153-8902, Japan}
\date{\today}
\maketitle

{\bf
Signal transduction in living cells is vital to maintain life itself, where information transfer in noisy environment plays a significant role.  In a rather different context, the recent intensive researches of ``Maxwell's demon" -- a feedback controller that utilizes information of individual molecules -- has led to a unified theory of information and thermodynamics. Here we combine these two streams of researches, and show that the second law of thermodynamics with information reveals the fundamental limit of the robustness of signal transduction against environmental fluctuations. Especially, we found that the degree of robustness is quantitatively characterized by an informational quantity called transfer entropy. Our information-thermodynamic approach is applicable to biological communication inside cells, in which there is no explicit channel coding in contrast to artificial communication. Our result would open up a novel biophysical approach to understand information processing in living systems on the basis of the fundamental information-thermodynamics link.}

	A crucial feature of biological signal transduction lies in the fact that it works in noisy environment~\cite{physbiolcell,Emmonet, Lestas}. To understand its mechanism, signal transduction has been modeled as noisy information processing~\cite{Andrews, SkerkerLaub, Mehta, TostevinWolde, Tu, CheongLevchenko,Kuroda, Govern}.  
For example, signal transduction of bacterial chemotaxis of {\it E. coli} ({\it Escherichia coli}) has been investigated as a simple model organism for sensory adaptation~\cite{Leibler, Leibler2, TuBerg, Shimizu, LanTu}. A crucial ingredient of {\it E. coli} chemotaxis is a feedback loop, which enhances the robustness of the signal transduction against environmental noise. 

The information transmission inside the feedback loop can be quantified by the transfer entropy, which was originally introduced in the context of time series analysis~\cite{Schreiber}, and has been studied in electrophysiological systems~\cite{Vicente}, chemical processes~\cite{Bauer}, and artificial sensorimotors~\cite{Lungerella}. The transfer entropy is the conditional mutual information representing the directed information flow, and gives an upper bound of the redundancy of the channel coding in an artificial communication channel with a feedback loop~\cite{Massey}; this is a fundamental consequence of Shannon's second theorem~\cite{Shannon, CoverThomas}. However, as there is not any explicit channel coding inside living cells, the role of the transfer entropy in biological communication has not been fully understood.

The transfer entropy also plays a significant role in thermodynamics~\cite{ItoSagawa}.
Historically, the connection between thermodynamics and information was first discussed in the thought experiment of ``Maxwell's demon'' in the nineteenth century~\cite{Maxwell, demon,Szilard}, where the demon is regarded as a feedback controller. In the recent progress on this problem in light of modern nonequilibrium statistical physics~\cite{Sekimoto, Seifert},  a universal and quantitative theory of thermodynamic feedback control has been developed, leading to the field of information thermodynamics~\cite{Allahverdyan, SagawaUeda2, Toyabe,HorowitzVaikuntanathan, Fujitani, HorowitzParrondo, ItoSano, SagawaUeda3, Kundu, MandalJarzynski, BerutLutz, ItoSagawa, Hartich, Munakata, HartichSeifert, Horowitz, SartoriHorowitz, LangMehta,HorowitzSandberg, Shiraishi}. Information thermodynamics reveals a generalization of the second law of thermodynamics, which implies that the entropy production of a target system is bounded by the transfer entropy from the target system to the outside world~\cite{ItoSagawa}.

 In this article, we apply the generalized second law to establish the quantitative relationship between the transfer entropy and the robustness of adaptive signal transduction against noise. We show that the transfer entropy gives the fundamental upper bound of the robustness, elucidating an analogy between information thermodynamics and the Shannon's information theory~\cite{Shannon, CoverThomas}. We numerically studied the information-thermodynamic efficiency of the signal transduction of {\it E. coli} chemotaxis, and found that the signal transduction of {\it E. coli} chemotaxis is efficient as an information-thermodynamic device, even when it is highly dissipative as a conventional heat engine.

{\bf Results}

{\bf Model.} 
The main components of {\it E. coli} chemotaxis are the ligand density change $l$, the kinase activity $a$, and the methylation level $m$ of the receptor (see also Fig.~1). A feedback loop exists between $a$ and $m$, which reduces the environmental noise in the signal transduction pathway from $l$ to $a$~\cite{SartoriTu}.
Let $l_t$, $a_t$, and $m_t$ be the values of these quantities at time $t$.  They obey stochastic dynamics due to the noise, and are described by the the following coupled Langevin equations~\cite{TostevinWolde, TuBerg, LanTu}:
\begin{equation}
\begin{split}
\dot{a}_t &=-\frac{1}{\tau^a} [a_t - \bar{a}_t (m_t, l_t) ]+\xi^a_t, \\
\dot{m}_t &=-\frac{1}{\tau^m} a_t+\xi^m_t,
\end{split}
\label{Langevin}
\end{equation}
where $\bar{a}_t(m_t, l_t) $ is the stationary value of the kinase activity under the instantaneous values of the methylation level $m_t$ and the ligand signal $l_t$. In the case of {\it E. coli} chemotaxis, we can approximate $\bar{a}_t(m_t, l_t) $ as $\alpha m_t -\beta l_t$, by linearizing it around the steady-state value~\cite{TostevinWolde, TuBerg}. $\xi^x_t$ ($x=a,m$) is the white Gaussian noise with $\langle \xi^x_t\rangle=0$ and $\langle \xi^x_t\xi^{x'}_{t'}\rangle=2T^x_t\delta_{xx'}\delta(t-t')$, where $\langle\cdots\rangle$ describes the ensemble average. $T^x_t$ describes the intensity of the environmental noise at time $t$, which is not necessarily thermal inside cells.  The noise intensity $T^a_t$ characterizes the ligand fluctuation. The time constants satisfy $\tau^m\gg\tau^a>0$, which implies that the relaxation of $a$ to $ \bar{a}_t$ is much faster than that of $m$.
					\begin{figure}
		\centering
		\includegraphics[width=85mm, bb=56 137 802 430]{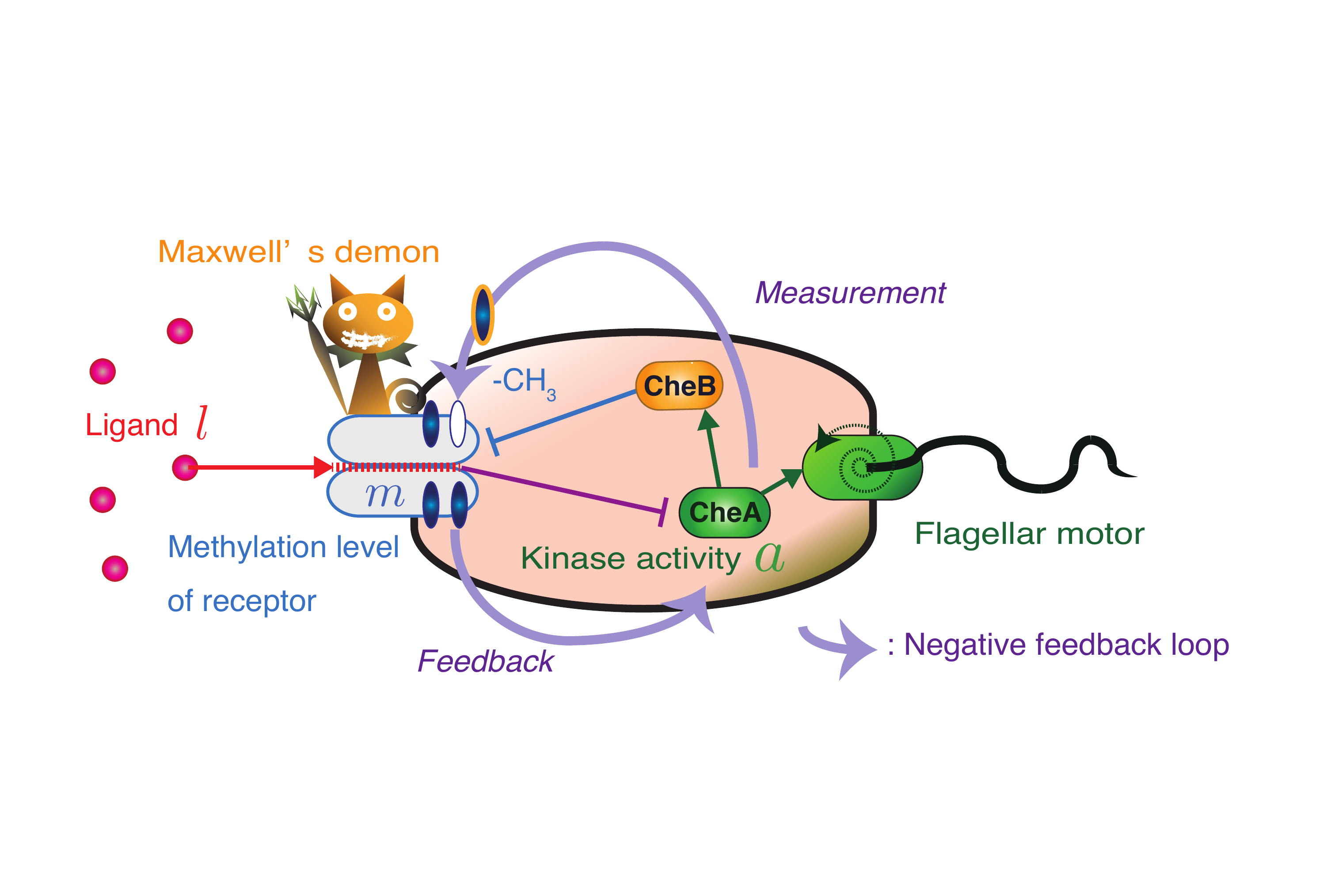}
		\caption{ {\bf Schematic of adaptive signal transduction of $\bm{E.}$  $\bm{coli}$ bacterial chemotaxis.}
Kinase activity $a$ (green) activates a flagellar motor to move {\it E. coli} toward a direction of the higher ligand density $l$ (red), by using the information stored in methylation level $m$ (blue). CheA is the histidine kinase related to the flagellar motor, and the response regulator CheB activated by CheA, removes methyl groups from the receptor. The methylation level $m$ plays a similar role to the memory of Maxwell's demon~\cite{Tu, ItoSagawa}, which reduces the effect of the environmental noise on the target system $a$; the negative feedback loop (purple arrows) counteracts the influence of ligand binding.}		\label{fig:chemotaxis}
	\end{figure}
	
The mechanism of adaptation in this model is as follows (see also Fig.~2)~\cite{TuBerg, LanTu}.
Suppose that the system is initially in a stationary state with $l_t = 0$ and $a_t =  \bar{a}_t  (m_t, 0)= 0$ at time $t<0$, and  $l_t$ suddenly changes from $0$ to  $1$ at time $t=0$ as a step function.
Then, $a_t$ rapidly equilibrates to $ \bar{a}_t (m_t, 1)$ so that the difference $a_t -  \bar{a}_t$ becomes small.
The difference $a_t -  \bar{a}_t$ plays an important role, which characterizes the level of adaptation.
Next, $m_t$ gradually changes to satisfy $ \bar{a}_t (m_t, 1) =0$, and thus $a_t $ returns to $0$, where $a_t -  \bar{a}_t$ remains small. 
						\begin{figure}
		\centering
		\includegraphics[width=85mm, bb=146 122 722 459]{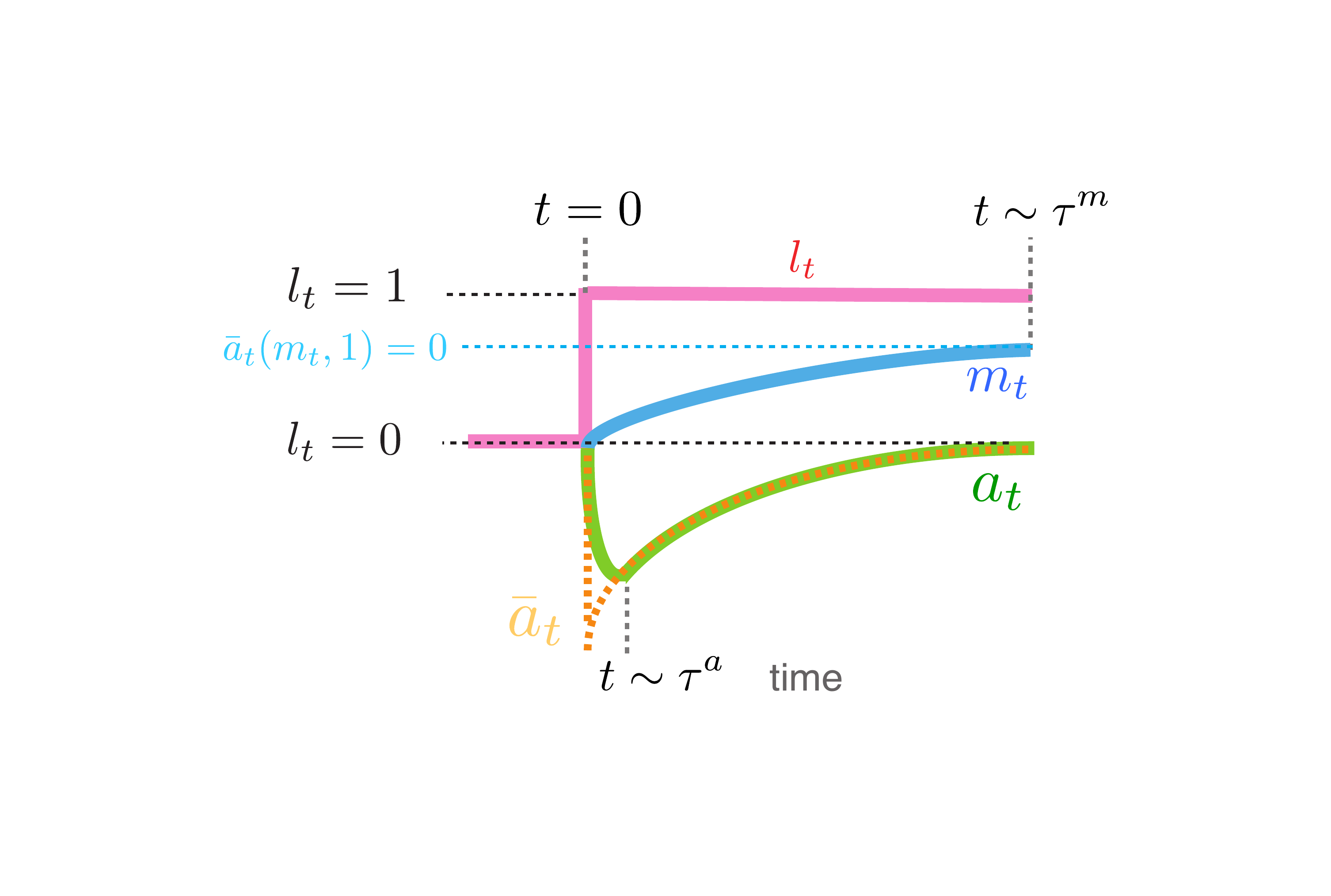}
		\caption{ {\bf Typical dynamics of adaptation with the ensemble average.} Suppose that  $l_t$ changes as a step function (red solid line). 
 Then, $a_t$ suddenly responds (green solid line), followed by the gradual response of $m_t$ (blue solid line).  
 The adaptation is achieved by the relaxation of $a_t$ to $ \bar{a}_t$ (orange dashed line). The methylation level $m_t$ gradually changes to $ \bar{a}_t (m_t, 1) =0$ (blue dashed line).}		\label{fig:chemotaxis}
	\end{figure}			

{\bf Robustness against environmental noise.} We introduce a key quantity that characterizes the robustness of adaptation, which is defined as the difference between the intensity of the ligand noise $T_t^a$ and the mean square error of the level of adaptation $\langle (a_t - \bar{a}_t)^2 \rangle$:
\begin{equation}
J^a_t  := \frac{1}{\tau^a } \left[  T^{a}_t - \frac{1}{\tau^a}\langle (a_t -  \bar{a}_t )^2 \rangle \right].
\label{equipartition}
\end{equation}
The larger $J^a_t$ is, the more robust the signal transduction is against the environmental noise. 
In the case of thermodynamics, $J^a_t$ corresponds to the heat absorption in $a$, and characterizes the violation of the fluctuation-dissipation theorem~\cite{Sekimoto}. Since the environmental noise is not necessarily thermal in the present situation, $J_t^a$ is not exactly the same as the heat, but is a biophysical quantity that characterizes the robustness of adaptation against the environmental noise. 

{\bf Information flow.} We here discuss the quantitative definition of the transfer entropy~\cite{Schreiber}. The transfer entropy from $a$ to $m$ at time $t$ is defined as the conditional mutual information between $a_t$ and $m_{t+dt}$ under the condition of $m_t$:
\begin{equation}
d I^{{\rm tr}}_t := \int dm_{t+dt} da_t dm_t p[m_{t+dt}, a_t, m_t] \ln \frac{p[m_{t+dt}| a_t, m_t ]}{p[m_{t+dt}| m_t] },
\label{TE}
\end{equation}
where $p[m_t, m_{t+dt}, a_t]$ is the joint probability distribution of $(a_t, m_t, m_{t+dt})$, and $p[m_{t+dt}|a_t,m_t]$ is the probability distribution of $m_{t+dt}$ under the condition of $(a_t, m_t)$.
The transfer entropy characterizes the directed information flow from $a$ to $m$ during an infinitesimal time interval $dt$~\cite{Schreiber, Schreiber2},  which quantifies a causal influence between them~\cite{SchindlerBhattacharya, BarnettSeth}. 
From the nonnegativity of the conditional mutual information~\cite{CoverThomas}, that of the transfer entropy follows: $d I^{{\rm tr}}_t \geq 0$.

{\bf Second law of information thermodynamics.} We now consider the second law of information thermodynamics, which characterizes the entropy change in a subsystem in terms of the information flow (see also Fig.~3). In the case of Eq. (\ref{Langevin}), the generalized second law is given as follows [see also Method]:
\begin{equation}
d I^{{\rm tr}}_t  + dS^{a|m}_t \geq \frac{J_t^a}{T^a_t} dt .
\label{bound}
\end{equation}
Here, $dS^{a|m}_t$ is the conditional Shannon entropy change defined as $dS^{a|m}_t:=S[a_{t+dt}|m_{t+dt}] - S[a_{t}|m_{t}]$ with $S[a_t|m_t] :=-\int{da_tdm_t}p[a_t,m_t]\ln{p}[a_t|m_t]$, which vanishes in the stationary state. The transfer entropy $dI_t^{\rm tr}$ on the left-hand side of (\ref{bound}) shows the significant role of the feedback loop, implying that the robustness of adaptation can be enhanced against the environmental noise by the feedback using information. This is analogous to the central feature of Maxwell's demon.
		\begin{figure}
		\centering
		\includegraphics[width=85mm, bb=25 99 826 476]{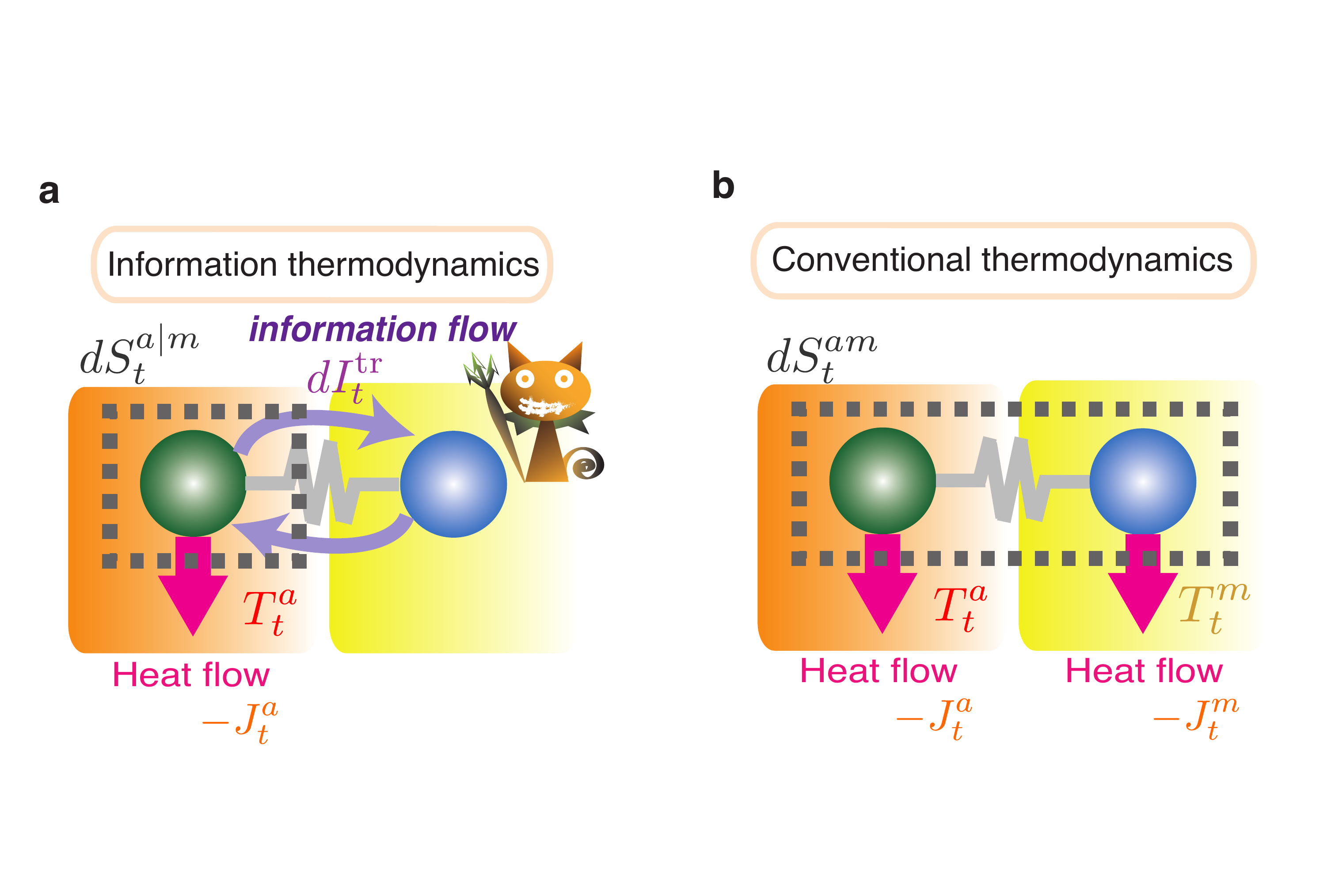}
		\caption{ {\bf Schematics of information thermodynamics and conventional thermodynamics.} A green (blue) circle indicates subsystem $a$ ($m$), and a gray polygonal line indicates their interaction. {\bf a}, The second law of information thermodynamics characterizes the entropy change in a subsystem in terms of the information flow between the subsystem and the outside world (i.e., $ \Xi^{\rm info}_t := d I^{{\rm tr}}_t  + dS^{a|m}_t \geq -J^a_t dt/ T^a_t$). The information-thermodynamic picture concerns the entropy change inside the dashed square that only includes subsystem $a$.
 {\bf b}, the conventional second law of thermodynamics states that the entropy change in a subsystem is compensated for by the entropy change in the outside world (i.e., $\Xi^{\rm SL}_t := -J^{m}_t dt/T^m_t+d S^{am}_t  \geq  J^a_t dt/T^a_t$). The conventional thermodynamic picture concerns the entropy change inside the dashed square, which includes the entire systems $a$ and $m$. As explicitly shown in this paper, information thermodynamics gives a tighter bound of the robustness $J^a_t$ in the biochemical signal transduction of {\it E. coli} chemotaxis.}		\label{fig:chemotaxis}
	\end{figure}

 To further clarify the meaning of inequality (\ref{bound}), we focus on the case of the stationary state.
If there was no feedback loop between $m$ and $a$, then the second law reduces to $\langle (a_t -  \bar{a}_t )^2 \rangle \geq\tau^aT^a_t$, which, as naturally expected, implies that the fluctuation of the signal transduction is bounded by the intensity of the environmental noise. 
In contrast, in the presence of a feedback loop, $\langle (a_t -  \bar{a}_t )^2 \rangle$ can be smaller than $\tau^aT^a_t$ owing to the transfer entropy $dI^{{\rm tr}}_t $ in the feedback loop:
\begin{equation}
\langle (a_t - \bar{a}_t )^2 \rangle  \geq \tau^a T^a_t \left[ 1  -\frac{dI^{{\rm tr}}_t }{dt} \tau^a \right].
\label{transfercooling}
\end{equation}
This inequality clarifies the role of the transfer entropy in biochemical signal transduction; the transfer entropy characterizes an upper bound of the robustness of the signal transduction in the biochemical network. The equality in (\ref{transfercooling}) is achieved in the limit of $\alpha \to 0$ and $\tau^a / \tau^m \to 0$ for the linear case with $\bar{a}_t(m_t, l_t) = \alpha m_t -\beta l_t$ (see Supplementary Note 1). The latter limit means that $a$ relaxes infinitely fast and the process is quasi-static (i.e., reversible) in terms of $a$. This is analogous to the fact that Maxwell's demon can achieve the maximum thermodynamic gain in reversible processes~\cite{HorowitzParrondo}. In general, the information-thermodynamic bound becomes tight if $\alpha$ and $\tau^m/\tau^a$ are both small.  The realistic parameters of the bacterial chemotaxis are given by $\alpha \simeq 3$ and $\tau^a / \tau^m \simeq 0.1$~\cite{TuBerg,LanTu, TostevinWolde}, and therefore the real adaptation process is accompanied by a finite amount of information-thermodynamic dissipation.

Our model of chemotaxis has the same mathematical structure as the feedback cooling of a colloidal particle by Maxwell's demon~\cite{ItoSano, Kundu, Munakata, HorowitzSandberg}, where the feedback cooling is analogous to the noise filtering in the sensory adaptation~\cite{SartoriTu}. This analogy is a central idea of our study; the information-thermodynamic inequalities [(\ref{transfercooling}) in our case] characterize the robustness of adaptation as well as the performance of feedback cooling.

{\bf Numerical result.}
We consider the second law (\ref{bound}) in non-stationary dynamics, and numerically demonstrate the power of this inequality. Figure 4 shows $J^a_t dt /T^a_t$ and
\begin{equation}
\Xi^{\rm{info}}_t := d I^{{\rm tr}}_t  + dS^{a|m}_t
\end{equation}
 in six different types of dynamics of adaptation, where the ligand signal is given by a step function (Fig.~4a), a sinusoidal function (Fig.~4b), a linear function (Fig.~4c), an exponential decay (Fig.~4d), a square wave (Fig.~4e), a triangle wave (Fig.~4f). These results confirm that  $\Xi^{\rm{info}}_t$ gives a tight bound of $J^a_t$, implying that the transfer entropy characterizes the robustness well. In Fig. 4b and 4f, the robustness $J^a_t dt/T^a_t$ is nearly equal to the information-thermodynamic bound $\Xi^{\rm info}_t$ when the signal and noise are decreasing or increasing rapidly (e.g., $t\simeq 0.008$ and $t=0.012$ in Fig. 4f).
 
	 			\begin{figure*}[tb]
		\centering
		\includegraphics[width=180mm, bb=25 189 1404 1135]{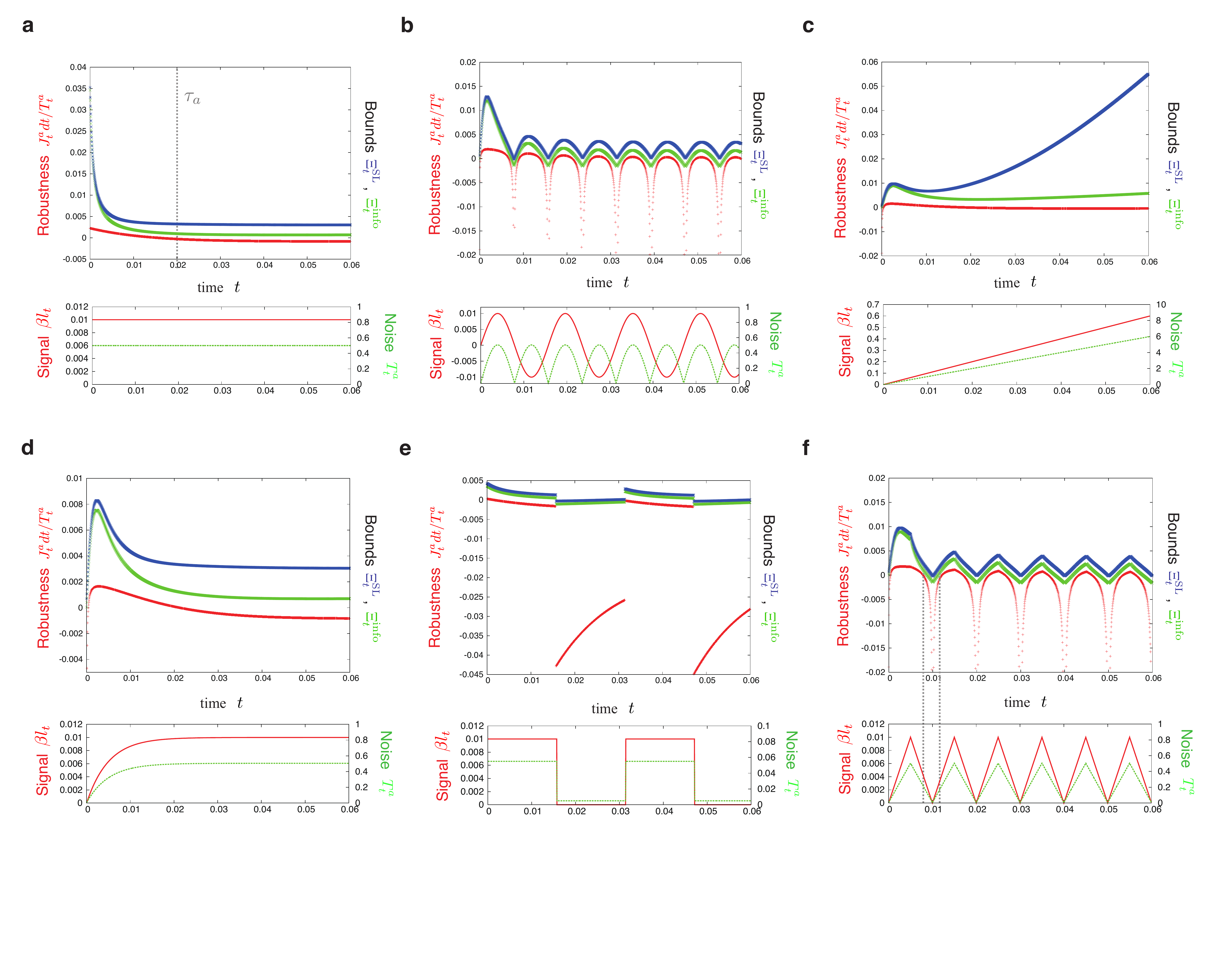}
		\caption{{\bf Numerical results of the information-thermodynamic bound on the robustness.}  
We compare the  robustness $J^a_t$ (red line), the information-thermodynamic bound $\Xi^{\rm info}_t$ (green line), and the conventional-thermodynamic bound  $\Xi^{\rm SL}_t$ (blue line).
The initial condition is the stationary state with $ \bar{a}_t= \alpha m_t -\beta l_t$, fixed ligand signal $\beta l_t = 0$, and noise intensity $T^a = 0.005$. We numerically confirmed that $\Xi^{\rm SL}_t \geq \Xi^{\rm info}_t \geq   J^{a}_t dt /T^{a}_t$ holds for the six transition processes. These results imply that,  for the signal transduction model, the information-thermodynamic bound is tighter than the conventional thermodynamic bound. The parameters are chosen as $\tau^a=0.02$, $\tau^m=0.2$, $\alpha=2.7$, and $T^{m}_t=0.005$, to be consistent with the real parameters of {\it E. coli} bacterial chemotaxis~\cite{TuBerg,LanTu, TostevinWolde}. We discuss the six different types of input signals $\beta l_t$ (red solid line) and noises $T^a_t$ (green dashed line). {\bf a}, Step function: $\beta l_t = 0.01$ and $T^a_t = 0.5$  for $t>0$. {\bf b}, Sinusoidal function: $\beta l_t = 0.01 \sin (400 t)$ and $T^a_t =0.5 |\sin (400 t) | +0.005$ for $t>0$. {\bf c}, Linear function: $\beta l_t = 10t$ and $T^a_t = 100t +0.005$ for $t>0$. {\bf d}, Exponential decay: $\beta L_t = 0.01 [1- \exp (-200t)]$ and $T^a_t = 0.5 [1- \exp (-200t)] +0.005$ for $t>0$. {\bf e}, Square wave: $\beta l_t = 0.01 [ 1+ \lfloor \sin (200 t) \rfloor]$ and $T^a_t = 0.05 [ 1+ \lfloor \sin (200 t) \rfloor] +0.005$ for $t>0$, where $\lfloor \dots \rfloor$ denotes the floor function.  {\bf f}, Triangle wave: $\beta l_t = 0.01 |2 ( 100 t - \lfloor 100 t + 0.5 \rfloor) |$ and $T^a_t = 0.5|2 ( 100 t - \lfloor 100 t + 0.5 \rfloor) |+0.005$ for $t>0$.}
		\label{fig:graph}
	\end{figure*}

{\bf Conventional second law of thermodynamics.}
For the purpose of comparison, we next consider another upper bound of the robustness, which is given by the conventional second law of thermodynamics without information. We define the heat absorption by $m$ as  $J^m_t:=-\langle{a_t^2}\rangle/(\tau^m)^2$, and the Shannon entropy change in the total system as $dS^{am}_t:=S[a_{t+dt},m_{t+dt}] - S[a_{t},m_{t}]$ with $S[a_t, m_t]:=-\int{da_tdm_t}p[a_t,m_t]\ln{p}[a_t, m_t]$, which vanishes in the stationary state. We can then show that
\begin{equation}
\Xi^{\rm{SL}}_t := -\frac{J^{m}_t }{T^m_t}dt+d S^{am}_t
\end{equation}
is an upper bound of ${J^{a}_t dt}/T^{a}_t$, as a straightforward consequence of the conventional second law of thermodynamics of the total system of $a$ and $m$~\cite{Sekimoto, Seifert}. 
The conventional second law implies that the dissipation in $m$ should compensate for that in $a$ [see also Fig. 3]. Figure 4 shows ${J^{a}_t dt}/T^{a}_t$ along with $\Xi^{\rm{info}}_t$ and $ \Xi^{\rm{SL}}_t$. Remarkably, information-thermodynamic bound $\Xi^{\rm{info}}_t$ gives a tighter bound of $J^a_t$ than the conventional thermodynamic bound $\Xi^{\rm{SL}}_t$ such that 
\begin{equation}
\Xi^{\rm{SL}}_t\geq\Xi^{\rm{info}}_t\geq \frac{J^{a}_t}{T^{a}_t}dt,
\label{twobound}
\end{equation}
 for every non-stationary dynamics shown in Fig.~4. Moreover, we can analytically show inequalities (\ref{twobound}) in the stationary state [see Supplementary Note 4].
 
To compare the information-thermodynamic bound and the conventional-thermodynamic one more quantitatively, we introduce an information-thermodynamic figure of merit based on inequalities (\ref{twobound}):
\begin{equation}
\chi :=1- \frac{\Xi^{\rm{info}}_t - J^{a}_t dt/T^{a}_t }{\Xi^{\rm{SL}}_t - J^{a}_t dt/T^{a}_t},
\end{equation}
where the second term on the right-hand side is given by the ratio between the information-thermodynamic dissipation $\Xi^{\rm{info}}_t - J^{a}_t dt/T^{a}_t$ and the entire thermodynamic dissipation $\Xi^{\rm{SL}}_t - J^{a}_t dt/T^{a}_t$. This quantity satisfies $0\leq \chi \leq 1$, and $\chi \simeq 1$ ($\chi \simeq 0$) means that information-thermodynamic bound is much tighter (a little tighter) compared to the conventional thermodynamic bound. We numerically calculated $\chi$ in the aforementioned six types of dynamics of adaptation [see Supplementary Fig. 1-6]. In the case of a linear function [Supplementary Fig. 3], we found that $\chi$ increases in time $t$ and approaches to $\chi \simeq 1$. In this case, the signal transduction of {\it E. coli} chemotaxis is highly dissipative as a thermodynamic engine, but efficient as an information transmission device.  

{\bf Comparison with Shannon's theory.} We here discuss the similarity and the difference between our result and the Shannon's information theory~\cite{Shannon, CoverThomas}  (see also Fig.~5). 
The Shannon's second theorem (i.e., the noisy-channel coding theorem) states that an upper bound of achievable information rate $R$ is given by the channel capacity $C$ such that $C\geq R$.
The channel capacity $C$ is defined as the maximum value of the mutual information with finite power, where the mutual information can be replaced by the transfer entropy $dI^{\rm tr}_t$ in the presence of a feedback loop~\cite{Massey}. $R$ describes how long bit sequence is needed for a channel coding to realize errorless communication through a noisy channel, where errorless means the coincidence between the input and output messages. Therefore, both of $J^a_t$ and $R$ characterize the robustness information transmission against noise, and bounded by the transfer entropy $dI^{\rm tr}_t$. In this sense, there exists an analogy between the second law of thermodynamics with information and the Shannon's second theorem. In the case of biochemical signal transduction, the information-thermodynamic approach is more relevant, because there is not any explicit channel coding inside cells.  Moreover, while $J^a_t$ is an experimentally measurable quantity as mentioned below~\cite{Sekimoto, Seifert}, $R$ cannot be properly defined in the absence of any artificial channel coding~\cite{CoverThomas}. Therefore, $J_t^a$ is an intrinsic quantity to characterize the robustness of the information transduction inside cells.
				\begin{figure}
		\centering
		\includegraphics[width=85mm, bb=30 19 706 710]{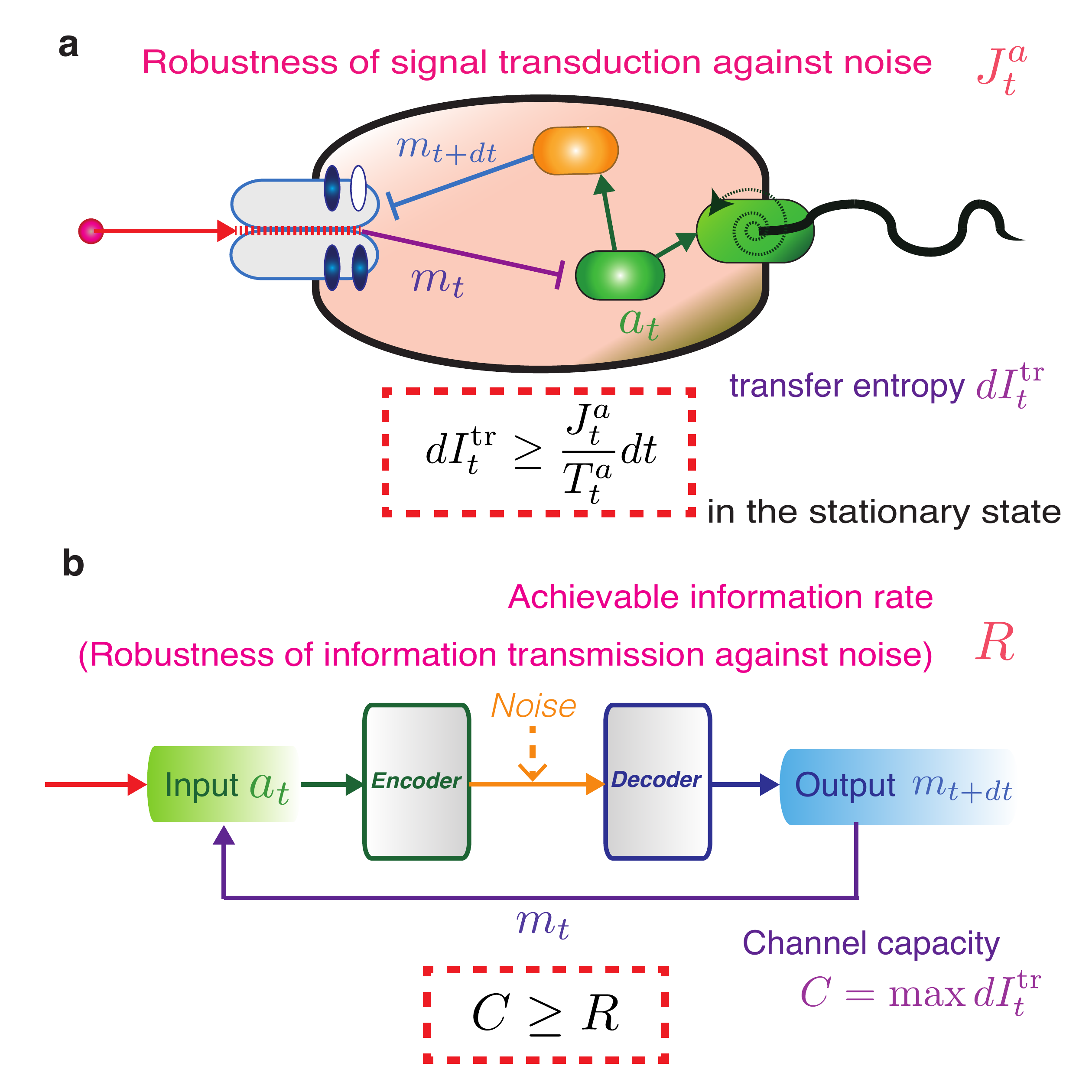}
		\caption{{\bf Analogy and difference between our approach and Shannon's information theory.} {\bf a}, Information thermodynamics for biochemical signal transduction. The robustness $J^a_t$ is bounded by the transfer entropy $dI^{\rm tr}_t$  in the stationary states, which is a consequence of the second law of information thermodynamics.   {\bf b}, Information theory for artificial communication.  The archivable information rate $R$, given by the redundancy of the channel coding, is bounded by the channel capacity $C =\max dI^{\rm tr}_t$, which is a consequence of the Shannon's second theorem. If the noise is Gaussian as is the case for the {\it E. coli} chemotaxis, both of the transfer entropy and the channel capacity are given by the power-to-noise ratio $C = d I^{\rm tr}_t= (2)^{-1} \ln (1+ d{P}_t /N_t )$, under the condition that the initial distribution is Gaussian [see Method].}
		\label{fig:discussion}
	\end{figure}

{\bf Discussion}

Our result can be experimentally validated, by measuring the transfer entropy and thermodynamic quantities from the probability distribution of the amount of proteins in a biochemical system [5, 6, 9, 10, 46-49].  In fact, the transfer entropy $dI_{\rm tr}$ and thermodynamic quantities (i.e., $dS^{a|m}_t$ and $J^a_t dt/T^a_t $) can be obtained from the joint probability distribution of $(a_t, m_t, a_{t+dt}, m_{t+dt})$. The measurement of such a joint distribution would not be far from today's experimental technique in biophysics~\cite{SkerkerLaub, Mehta, CheongLevchenko,Kuroda,Collin,Ritort,Toyabe2, Hayashi}. Experimental measurements of $\Xi^{\rm info}_t$ and $J^a_t dt/T^a_t$ would lead to a novel classification of signal transduction in terms of the thermodynamic cost of information transmission.

We note that, in Ref.~\cite{LanTu}, the authors discussed that the entropy changes in two heat baths $-J^a_t/T_t^a- J^m_t/T_t^m \simeq- J^m_t/T_t^m=\langle{a_t^2}\rangle/[T_t^m (\tau^m)^2]$ can be characterized by the accuracy of adaptation. In our study, we derived a bound for $J^a_t dt/T^a_t$ that is regarded as the robustness of signal transduction against the environmental noise. These two results capture complementary aspects of adaptation processes: accuracy and robustness.

We also note that our theory of information thermodynamics~\cite{ItoSagawa} can be generalized to a broad class of signal transduction networks, including a feedback loop with time delay.

{\bf Method}

{\bf The outline of the derivation of inequality (\ref{bound}).}  
We here show the outline of the derivation of the information-thermodynamic inequality (\ref{bound})  [see also Supplementary Note 2 for details]. The heat dissipation $J^a_t dt/ T^a_t$ is given by the ratio between forward and backward path probabilities as $J^a_t dt/ T^a_t =  \int da_t da_{t+dt} dm_t p [a_{t}, a_{t+dt},m_t]  \ln [ p_B [a_{t}|a_{t+dt},m_t] /p[a_{t+dt}|a_t,m_t] ]$~\cite{Sekimoto, Seifert, ItoSagawa}, where the backward path probability $p_B [a_{t}|a_{t+dt},m_t] := \mathcal{G} (a_t; a_{t+dt};m_t)$ can be calculated from the forward path probability $p[a_{t+dt}|a_t,m_t] =: \mathcal{G} (a_{t+dt}; a_{t};m_t)$. Thus, the difference $dI^{\rm tr}_t + dS^{a|m}_t - J^a_t dt/ T^a_t$ is given by the Kullback-Libler divergence~\cite{CoverThomas}. From its nonnegativity~\cite{CoverThomas}, we have $dI^{\rm tr}_t + dS^{a|m}_t \geq J^a_t dt/ T^a_t$. This inequality can be derived from the general inequality of information thermodynamics~\cite{ItoSagawa}. As discussed in Supplementary Note 3, this inequality gives a weaker bound of the entropy production. 

{\bf The analytical expression of the transfer entropy.} 
In the case of {\it E. coli} chemotaxis, we have $ \bar{a}_t= \alpha m_t -\beta l_t$, and Eqs.~(\ref{Langevin}) become linear. In this situation, if the initial distribution is Gaussian, we analytically obtain the transfer entropy up to the order of $dt$ [see also Supplementary Note 4]: $d I^{\rm tr}_t= (2)^{-1} \ln (1+ d{P}_t /N_t )$, where $N_t:=2T^m_t$ describes the intensity of the environmental noise, and $dP_t :=   [1- (\rho_t^{am})^2 ] V_t^a dt /(\tau^m)^2$ describes the intensity of the signal from $a$ to $m$ per unit time with $V^x_t := \langle x^2_t \rangle - \langle x_t \rangle^2$, and $\rho_t^{am} :=  [\langle a_t m_t \rangle -\langle a_t \rangle \langle m_t \rangle] / (V^a_t V^m_t)^{1/2}$.  We note that $dI_{\rm{tr}}$ for the Gaussian case is greater than that of the non-Gaussian case, if $V^x_t$ and $\rho_t^{am}$ are the same~\cite{CoverThomas}.
We also note that the above analytical expression of $dI^{\rm tr}_t$ is the same form as the Shannon-Hartley theorem~\cite{CoverThomas}.

\subsection{Acknowledgements}
	We are grateful to S-I. Sasa, U. Seifert, M. L. Rosinberg, N. Shiraishi, K. Kawaguchi, H. Tajima, A. C. Barato, D. Hartich, and M. Sano for valuable discussions. This work was supported by the Grants-in-Aid for JSPS Fellows (Grant No. 24$\cdot$8593), by JSPS KAKENHI Grant No. 25800217 and No. 22340114, by KAKENHI No. 25103003 ``Fluctuation \& Structure'',
 and by Platform for Dynamic Approaches to Living System from MEXT, Japan.
\subsection{Author contributions}	
S.I. mainly constructed the theory, carried out the analytical and numerical calculations, and wrote the paper. T.S. also constructed the theory, and wrote the paper. Both authors discussed the results at the all stages.
\subsection{Additional information}
The authors declare no competing financial interests. Supplementary information accompanies this paper.
\newpage

\newpage
\appendix
\section{SUPPLEMENTARY INFORMATION}

{\bf Supplementary note 1 $\mid$ Explicit expression of the information-thermodynamic dissipation.}

We consider the coupled Langevin equations (2) in the main text,
\begin{align}
\dot{a}_t &=-\frac{1}{\tau^a} [a_t -  \bar{a}_t (m_t, l_t) ]+\xi^a_t, 
\label{sup:Langa}\\
\dot{m}_t &=-\frac{1}{\tau^m} a_t+\xi^m_t,
\label{sup:Langb}
\end{align}
where $\xi^x_t$ ($x=a, m$) is a white Gaussian noise with the variance $2 T^x_t$: $\langle \xi^x_t \rangle = 0 $, and $\langle \xi^x_t  \xi^{x'}_{t'}  \rangle = 2 T^x_t \delta_{xx'} \delta(t-t')$.  In the model of E. coli bacterial chemotaxis given by Eqs. (\ref{sup:Langa}) and (\ref{sup:Langb}) with $ \bar{a}_t = \alpha m_t - \beta l_t$, we can analytically calculate the information-thermodynamic dissipation in the stationary state:
 \begin{align}
dI^{\rm tr}_t - \frac{J^a_t}{T^a_t}  dt&=  \frac{[\langle a_t^2 \rangle-\langle a_t \rangle^2][1-(\rho_t^{am} )^2 ]dt}{ 4(\tau^m )^2 T_t^m}+\frac{dt}{\tau^a T_t^a} \left[ \frac{1}{\tau^a} \langle (a_t-\bar{a}_t))^2 \rangle-T_t^a \right].
\label{sup:dissip}
\end{align}
When this quantity becomes zero, the equality in inequality (5) in the main text is achieved.  With the linear approximation $ \bar{a}_t = \alpha m_t - \beta l_t$, we can explicitly calculate the stationary values of $\langle a_t \rangle$, $\langle m_t \rangle$, $\langle a_t^2 \rangle$, $\langle a_t m_t \rangle$ and $\langle m_t^2 \rangle$ as
\begin{align}
\langle a_t \rangle_{\rm SS} &=0,\\
\langle m_t \rangle_{\rm SS} &=\beta \alpha^{-1} l_t, \\
\langle a_t^2 \rangle_{\rm SS} &= \alpha \tau^m T_t^m+\tau^a T_t^a,\\
\langle a_t m_t \rangle_{\rm SS} &=\tau^m T_t^m,\\
\langle m_t^2 \rangle_{\rm SS} &=(\beta \alpha^{-1} l_t )^2+\alpha^{-1} \tau^m T_t^m+ \tau^a \alpha^{-1} (\tau^m)^{-1} [\alpha \tau^m T_t^m+\tau^a T_t^a].
\end{align}
The information-thermodynamic dissipation (\ref{sup:dissip}) then reduces to
 \begin{align}
dI^{\rm tr}_t - \frac{J^a_t}{T^a_t}  dt &=  dt[\alpha T_t^m+\tau^a (\tau^m)^{-1} T_t^a ] \left[\frac{\alpha}{\tau^a T_t^a } +\frac{1-(\rho_t^{am})^2 }{4\tau^m T_t^m} \right]\\
&\geq 0.
\end{align}
where the correlation coefficient $(\rho_t^{am} )^2$ is given by
 \begin{align}
(\rho_t^{am} )^2 &=\frac{1}{\left[ 1+ \tau^a (\tau^m)^{-1}[\alpha+\tau^a T_t^a (\tau^m T_t^m)^{-1}] \right] [1+ \tau^a T^a_t (\alpha \tau^m T^m_t)^{-1}] }  \\
 &\leq 1.
\end{align} 
In the limit of $\alpha \to 0$ and $\tau^a /\tau^m \to 0$, the information-thermodynamic dissipation (\ref{sup:dissip}) can be zero, and the equality in Eq. (5) in the main text is achieved such that
 \begin{align}
dI^{\rm tr}_t = \frac{J^a_t}{T^a_t}  dt =0.
\end{align} 
This corresponds to the situation where the feedback loop does not work ($\alpha \to 0$) and the information flow vanishes, and $a$ relaxes infinitely fast ($\tau^a /\tau^m \to 0$).

　

　

{\bf Supplementary note 2 $\mid$ Detailed derivation of the second law of information thermodynamics}

Here, we show the detailed derivation of the second law of information thermodynamics for Eqs. (\ref{sup:Langa}) and (\ref{sup:Langb}) [Eq. (4) in the main text]:
 \begin{align}
\Xi_t^{\rm info} := dI^{\rm tr}_t  + dS^{a|m}_t \geq \frac{J^a_t}{T^a_t} dt,
\label{sup:infothermo}
\end{align}
where $dS^{a|m}_t:=S[a_{t+dt}|m_{t+dt}] - S[a_{t}|m_{t}]$ is the conditional Shannon entropy change of $a$ with $S[a_t|m_t] :=- \int da_t dm_t p [a_t,m_t] \ln{p} [a_t|m_t]$, and $d I^{{\rm tr}}_t$ is the transfer entropy from $a$ to $m$ at time $t$:
\begin{equation}
d I^{{\rm tr}}_t := \int dm_{t+dt} da_t dm_t p[m_{t+dt}, a_t, m_t] \ln \frac{p[m_{t+dt}| a_t, m_t ]}{p[m_{t+dt}| m_t] }.
\label{sup:transfer}
\end{equation}
The heat absorption~\cite{sup:Sekimoto} $J^a_t$ is defined as the ensemble average of the Stratonovich product of the force $\xi^a_t -\dot{a}_t $ and the velocity $\dot{a}_t$ such that
\begin{equation}
J^a_t := \langle [ \xi^a_t -\dot{a}_t] \circ \dot{a}_t \rangle.
\end{equation}
The heat absorption $J_t^a$ can be rewritten by Eq. (3) in the main text: 
\begin{align}
J^a_t &= \langle [ \xi^a_t -\dot{a}_t] \circ \dot{a}_t  \rangle \nonumber\\
&=\frac{1}{\tau^a} \left[ \langle[ a_t -  \bar{a}_t] \circ \xi^a_t  \rangle -\frac{1}{\tau^a} \langle ( a_t - \bar{a}_t )^2 \rangle \right]  \nonumber\\
&=\frac{1}{\tau^a } \left[  T^{a}_t  -\frac{1}{\tau^a} \langle ( a_t - \bar{a}_t )^2 \rangle \right],
\end{align}
where we used the relation of the Stratonovich integral~\cite{sup:Sekimoto} $\langle f(a_t,m_t, l_t) \circ \xi^a_t \rangle = T^a_t \langle \partial_{a_t} f(a_t, m_t , l_t) \rangle$ for any function $f$. 

From the detailed fluctuation theorem~\cite{sup:Seifert}, $J^a_t dt/ T^a_t$ can be rewritten as a ratio of the probability distribution. Let the backward path-probability $p_B [a_{t}|a_{t+dt},m_t]$ be $p_B [a_{t}|a_{t+dt},m_t] := \mathcal{G} (a_t ; a_{t+dt}  ; m_t)$, where $\mathcal{G}$ is given by the path-integral expression:
\begin{align}
p [a_{t+dt}|a_t,m_t] &= \mathcal{N} \exp \left[ - \frac{dt}{4 T^a_t} \left[ \frac{a_{t+dt} - a_t}{ dt} +\frac{1}{\tau^a} [a_t -  \bar{a}_t (m_t, l_t) ]\right]^2 \right] \\
&=: \mathcal{G} (a_{t+dt} ; a_t ; m_t).
\end{align}
$\mathcal{N}$ is the normalization constant, so that $\int da_{t+dt} \mathcal{G} (a_{t+dt} ; a_t ; m_t) =1$ is satisfied. The backward path probability also satisfies the normalization condition $\int da_t p_B [a_{t}|a_{t+dt},m_t]= \int da_t \mathcal{G} (a_t ; a_{t+dt}  ; m_t) =1$. Up to order $dt$, the entropy change in the heat bath with temperature $T^a_t$ is calculated as
\begin{align}
\frac{J^a_t}{T^a_t} dt= \int da_t dm_t da_{t+dt} p[a_t , m_t, a_{t+dt}] \ln \frac{p_B [a_{t}|a_{t+dt},m_t]}{p [a_{t+dt}|a_t,m_t]},
\label{sup:detailed}
\end{align}
which is well known as the detailed fluctuation theorem~\cite{sup:Seifert}.

Because of the noise independence $\langle \xi^a_t  \xi^{m}_{t'}  \rangle = 0$, we have $p [a_{t+dt}, m_{t+dt},a_t, m_t] = p [a_{t+dt} | a_t, m_t] p [m_{t+dt} |a_t, m_t ] p[a_t, m_t]$. From Eqs.  (\ref{sup:transfer}) and (\ref{sup:detailed}), the difference $\Xi_t^{\rm info} - J^a_t dt/ T^a_t$ is calculated as
 \begin{align}
\Xi_t^{\rm info} - \frac{J^a_t}{T^a_t}  dt&=  \left< \ln \frac{p [ a_t, m_t, a_{t+dt} , m_{t+dt} ] }{p [a_{t+dt} | m_{t+dt}] p_B [a_{t}|a_{t+dt},m_t ] p [m_{t+dt},m_t] } \right>.
\end{align}
The quantity $\mathcal{Q} [a_{t},m_t, a_{t+dt},m_{t+dt}] := p [a_{t+dt} | m_{t+dt}] p_B [a_{t}|a_{t+dt},m_t ] p [m_{t+dt},m_t]$ satisfies the normalization condition of the probability:
\begin{align}
\int  da_t dm_t da_{t+dt}  dm_{t+dt}   \mathcal{Q} [a_{t},m_t, a_{t+dt},m_{t+dt}]=1. 
\end{align}

Therefore, $\mathcal{Q}  [a_{t},m_t, a_{t+dt},m_{t+dt}]$ can be interpreted as the probability distribution of $(a_t, m_t, a_{t+dt}, m_{t+dt})$, and the difference $\Xi_t^{\rm info} - J^a_t dt/ T^a_t$  is rewritten as the Kullback-Libler divergence $D_{\rm KL}(p|| \mathcal{Q})$~\cite{sup:CoverThomas}:
\begin{align}
\Xi_t^{\rm info} - J^a_t dt/ T^a_t  &=  \int da_t dm_t da_{t+dt} dm_{t+dt} p[a_{t},m_t, a_{t+dt},m_{t+dt}] \ln \frac{ p[ a_{t},m_t, a_{t+dt},m_{t+dt}]}{\mathcal{Q} [a_{t},m_t, a_{t+dt},m_{t+dt}]}
\\
&:=D_{\rm KL}(p|| \mathcal{Q}).
\end{align}
From the non-negativity of the Kullback-Leibler divergence~\cite{sup:CoverThomas} [i.e., $D_{\rm KL}(p|| \mathcal{Q}) \geq 0$], we obtain Eq.~(\ref{sup:infothermo}).

　

　

{\bf Supplementary note 3 $\mid$ Relationship between information thermodynamics for two-dimensional Markov process and that in [S. Ito and T. Sagawa, Phys. Rev. Lett. \bf{111}, 180503 (2013)]}

In our previous paper~\cite{sup:ItoSagawa}, we have derived a general framework of information thermodynamics and discussed information thermodynamics for the coupled Langevin equations. We here give another application of the general result in Ref.~\cite{sup:ItoSagawa} to two-dimensional Markov processes such as the coupled Langevin equations ~(\ref{sup:Langa}) and (\ref{sup:Langb}). Here, we show that the general result  in Ref.~\cite{sup:ItoSagawa} is tighter than the information-thermodynamic inequality (\ref{sup:infothermo}).

We first consider the path probability of a single time step from $\{ a_t, m_t \}$ to $\{ a_{t+dt}, m_{t+dt} \}$. Due to the Markov property, the joint probability $p [a_{t+dt}, m_{t+dt} ,a_{t}, m_{t}]$ is given by
\begin{align}
p [a_{t+dt}, m_{t+dt} ,a_{t}, m_{t} ] =  p[a_{t}, m_{t}] p[a_{t+dt}| a_{t}, m_{t} ]p[m_{t+dt}| a_{t}, m_{t} ],
\label{sup:oneprocess}
\end{align}
where the independency of the noise (i.e., $p[a_{t+dt} , m_{t+dt}| a_{t}, m_{t} ] = p[a_{t+dt}| a_{t}, m_{t} ]p[m_{t+dt}| a_{t}, m_{t} ]$) is assumed.

We next consider a Bayesian network which represents the stochastic process of Eq.~(\ref{sup:oneprocess}) (see Supplementary Figure 7). This Bayesian network is given by the parents (denoted as ``$\rm pa$") of the random variables: ${\rm pa}(a_{t})= m_{t}$, ${\rm pa}(m_{t})= \emptyset$,  ${\rm pa}(a_{t+dt})= \{ a_{t},  m_{t}\}$ and ${\rm pa}(a_{t+dt})= \{ a_{t}, m_{t}\}$. The stochastic process of Eq.~(\ref{sup:oneprocess}) is given by $p[a_{t+dt} , m_{t+dt}, a_{t}, m_{t} ]  = p [a_{t+dt}|{\rm pa}(a_{t+dt})]p [m_{t+dt}|{\rm pa}( m_{t+dt})]p [a_{t}|{\rm pa}(a_{t})] p[ m_{t}|{\rm pa}( m_{t})]$. This Bayesian network shows a single time step of the Markovian dynamics from time $t$ to time $t+dt$.

Let stochastic mutual information be $I [\mathcal{A}_1:\mathcal{A}_2] := \ln p [\mathcal{A}_1,\mathcal{A}_2 ] - \ln p [\mathcal{A}_1] - \ln p [\mathcal{A}_2] $, and stochastic conditional mutual information be $I [\mathcal{A}_1:\mathcal{A}_2|\mathcal{A}_3]  := \ln p [\mathcal{A}_1,\mathcal{A}_2 |\mathcal{A}_3] - \ln p [\mathcal{A}_1|\mathcal{A}_3 ] - \ln p [\mathcal{A}_2|\mathcal{A}_3] $, where $\mathcal{A}_1$, $\mathcal{A}_2$ and $\mathcal{A}_3$ are any set of random variables. From the argument in Ref.~\cite{sup:ItoSagawa}, the bound of the entropy production for the subsystem $a$ is given by an informational quantity $\Theta$, which corresponds to the Bayesian network shown in Supplementary Figure 7:
 \begin{align}
 \Theta &:= I_{\rm fin} -I_{\rm ini} -\sum_{l=1}^2 I_{\rm tr}^l,\\
 I_{\rm ini} & = I[a_{t}: {\rm pa}(a_{t})] \nonumber\\
 &= I [a_{t}: m_{t} ],\\
 I_{\rm tr}^1 &= I [c_1  :  {\rm pa}_X(c_1)] \nonumber\\
 &= 0,\\
  I_{\rm tr}^2 &= I [c_2:  {\rm pa}_X(c_2)|c_1] \nonumber\\
 &=I[ a_{t}: m_{t+dt}|m_{t} ],\\
 I_{\rm fin} &:= I [x_2 : \mathcal{C}] \nonumber\\
 &=I [a_{t+dt}: \{m_{t} , m_{t+dt}\} ],
 \end{align}
where we set $X := \{x_1 = a_t,  x_2 =a_{t+dt} \}$, $\mathcal{C} := \{c_1 = m_{t}, c_2 = m_{t+dt}\}$,  ${\rm pa}_X(m_{t}) := {\rm pa}(m_{t}) \bigcap  X =\emptyset$, and ${\rm pa}_X(m_{t+dt}) := {\rm pa}(m_{t+dt}) \bigcap  X =a_{t}$. Let the entropy production in the subsystem during the infinitesimal time step be $\sigma_t := \ln p [a_{t} ] -\ln p [a_{t+dt}] + \Delta s^{\rm bath}_t$, where $\Delta s^{\rm bath}_t$ is the entropy change in the heat baths. Again from the argument in Ref.~\cite{sup:ItoSagawa}, we have inequality $\langle \sigma_t \rangle \geq \langle \Theta \rangle$, where
 \begin{align}
\langle \Theta \rangle =& \langle I [a_{t+dt}:  \{ m_{t} , m_{t+dt}\}  ] \rangle -\langle I [a_{t}: \{ m_{t} , m_{t+dt}\} ] \rangle
\label{sup:infoflow}\\
 =& I_{t+dt}^{am}  - I_t^{am}  +d I^{\rm Btr}_{t}  - d I^{\rm tr}_t ,
 \label{sup:informationthermodynamics}
 \end{align} 
$I^{am}_t := \langle I [ a_t : m_t] \rangle$ is the mutual information between $a$ and $m$ at time $t$,  $d I^{\rm tr}_{t}  := \langle\ln p[ m_{t+dt}|a_{t}, m_{t} ]\rangle  -  \langle\ln p[ m_{t+dt}|m_{t} ] \rangle$ is the transfer entropy from $a$ to $m$ at time $t$, and $d I^{\rm Btr}_{t}$ is defined as the conditional mutual information $d I^{\rm Btr}_{t}  := \langle\ln p [m_{t}|m_{t+dt} , a_{t+dt}]\rangle  -  \langle\ln p [m_{t}|m_{t+dt}] \rangle$. We note Eq. (\ref{sup:infoflow}) is consistent with information flow in several papers~\cite{sup:Allahverdyan,sup:Hartich,sup:Horowitz,sup:Shiraishi}.
 
For the two-dimensional Langevin system Eqs.~(\ref{sup:Langa}) and (\ref{sup:Langb}), the ensemble average of the entropy production for the subsystem $\langle \sigma_t \rangle$ can be rewritten by the heat absorption $J^a_t$, $\langle \sigma_t \rangle =-J^a_t dt /T_t^a +\langle \ln p[a_t] \rangle - \langle \ln p[ a_{t+dt}] \rangle$ with $\langle \Delta s^{\rm bath}_t \rangle =-J^a_t dt /T_t^a $. From  $\langle \sigma_t \rangle \geq \langle \Theta \rangle$, we have the following inequality:
\begin{align}
-d I^{\rm Btr}_{t}  + d I^{\rm tr}_t +  dS^{a|m}_t &\geq \frac{J^a_t}{T_t^a}  dt.
\end{align}
where we used Eq.~(\ref{sup:informationthermodynamics}) and identity $dS^{a|m}_t = \langle \ln p[a_t] \rangle - \langle \ln p[ a_{t+dt}]  \rangle -I_{t+dt}^{am}  + I_t^{am}$. Because of  the non-negativity of the mutual information~\cite{sup:CoverThomas} [i.e., $dI^{\rm Btr}_t \geq 0$], we have inequality (\ref{sup:infothermo})  [Eq. (4) in the main text]:
\begin{align}
\frac{J^a_t}{T_t^a} dt &\leq-d I^{\rm Btr}_{t}  + d I^{\rm tr}_t +  dS^{a|m}_t  \label{sup:ineq1}\\
&\leq  d I^{\rm tr}_t +  dS^{a|m}_t.
 \label{sup:ineq2}
 \end{align} 
The conditional mutual information $d I^{\rm Btr}_t$ would be important as well as the transfer entropy $dI^{\rm tr}_t$, because the bound including $d I^{\rm Btr}_t$ [Eq.~(\ref{sup:ineq1})] is tighter than the bound without $d I^{\rm Btr}_t$ [Eq.~(\ref{sup:ineq2})]. However, in the main text, we only focus on the role of the transfer entropy $dI^{\rm tr}_t$ for the sake of simplicity, by applying the weaker inequality~(\ref{sup:ineq2}). 

　

　

\section{Supplementary note 4 $\mid$ Analytical calculation of the transfer entropy for the coupled linear Langevin system}

We derive the analytical expression of the transfer entropy for the coupled linear Langevin system: 
\begin{align}
\dot{x}^1_t &= \sum_j \mu^{1j}_t x_t^j +f^1_t +\xi^1_t, \nonumber\\
\dot{x}^2_t &= \sum_j \mu^{2j}_t x_t^j +f^2_t +\xi^2_t, \nonumber\\
\langle \xi^i_t \xi^j_{t'} \rangle &=2 T^i_t \delta_{ij} \delta(t-t') \nonumber\\
\langle \xi^i_t\rangle &=0,
\label{sup:Langevin3d3}
\end{align}
where $i,j=1, 2$, $f^i_t$ and $\mu^{ij}_t$ are the time-dependent constants, $T_t^i$ is time-dependent variance of the white Gaussian noise $\xi^i_t$, and $\langle \dots \rangle$ denotes the ensemble average. In the main text, we considered the model of the {\it E. coli} bacterial chemotaxis given by Eqs.~(\ref{sup:Langa}) and (\ref{sup:Langb}) with $ \bar{a}_t = \alpha m_t - \beta l_t$. To compare Eqs.~(\ref{sup:Langa}) and (\ref{sup:Langb}), we set $\{x^1_t, x^2_t\} = \{ a_t, m_t\}$, $\mu^{11}_t = - 1/\tau^a$, $\mu^{12}_t = \alpha / \tau^a$, $f^1_t = - \beta l_t/ \tau^a$, $\mu^{21}_t = -1/\tau^m$, $\mu^{22}_t = 0$, $f^2_t =0$, $T^1_t =T^a_t$, and $T^2_t =T^m_t$. The transfer entropy from the target system $x^1$ to the other system $x^2$ at time $t$ is defined as $d I^{\rm tr}_t  := \langle \ln p [ x^{2}_{t+dt}| x^1_t, x_t^{2} ]\rangle - \langle \ln p[ x^{2}_{t+dt}| x_{t}^{2}] \rangle$.

Here, we analytically calculate the transfer entropy for the case that the joint probability $p[x^1_t, x_t^{2}]$ is a Gaussian distribution:
\begin{equation}
p[x^1_t, x_t^{2}] = \frac{1}{(2\pi) \sqrt{\det{\Sigma_t} }} \exp \left[ - \sum_{ij} \frac{1}{2} \bar{x}^i_t G^{ij}_t \bar{x}^j_t \right],
\label{sup:Gaussian}
\end{equation}
where $\Sigma^{ij}_t$ is the covariant matrix $\Sigma^{ij}_t := \langle  x^i_t x^j_t\rangle -\langle x^i_t \rangle \langle x^j_t\rangle$, and $\bar{x}^j_t := x^j_t - \langle x^j_t \rangle$. The inverse matrix $G_t := \Sigma_t^{-1}$ satisfies $ \sum_j G^{ij}_t\Sigma^{jl}_t = \delta_{il}$ and $G^{ij}_t = G^{ji}_t$. The joint distribution $p [x^2_t]$ is given by the Gaussian probability:
\begin{align}
p[ x^2_t ]&=\frac{1}{\sqrt{ 2\pi \Sigma_t^{22}}} \exp \left[ - \frac{1}{2} (\Sigma^{22}_t)^{-1} (\bar{x}^{2}_t)^2 \right].
\label{sup:Gaussian2}
\end{align}

We consider the path-integral expression of the Langevin equation (\ref{sup:Langevin3d3}). The conditional probability $p [ x^2_{t+dt}| x^1_t, x^{2}_t ]$ is given by
\begin{align}
p[ x^2_{t+dt}| x^1_t, x_t^2 ] &= {\mathcal N} \exp \left[ - \frac{dt}{4T^2_t} \left( \frac{x^2_{t+dt} - x^2_t}{dt} - \sum_j \mu^{2j}_t x_t^j -f^2_t \right)^2 \right]\\
&= {\mathcal N} \exp \left[ - \frac{dt}{4T^2_t} \left( F^2_t-\mu^{21}_t \bar{x}^1_t  \right)^2 \right]
\label{sup:pathintegral}
\end{align}
where ${\mathcal N}$ is the normalization constant with $\int dx^2_{t+dt} p[ x^2_{t+dt}| x^1_t, x_t^2 ] =1$.
For the simplicity of notation, we set $F^2_t =  (x^2_{t+dt} - x^2_t) /dt- \mu^{21}_t \langle x_t^1 \rangle -  \mu^{22}_t x_t^2  -f^2_t $. From Eqs. (\ref{sup:Gaussian}) and (\ref{sup:pathintegral}), we have the joint distribution $p [x^{2}_{t+dt}, x^{2}_t]$ as
\begin{align}
p [x^{2}_{t+dt}, x^{2}_t] =&\int dx^1_t  p[ x^2_{t+dt}| x^1_t, x_t^2 ]  p[x^1_t, x_t^{2}] \nonumber \\
=&\frac{{\mathcal N} }{ \sqrt{ 4\pi \det{\Sigma_t }\left( \frac{d t}{4 T^2_t} (\mu^{2 1}_t )^2 +\frac{G^{11}_t}{2}\right)}}  \exp \left[ - \frac{d t}{4T^2_t} (F^2_t)^2  - \frac{1}{2} G^{22}_t (\bar{x}^{2}_t)^2  + \frac{\left(  G^{12}_t \bar{x}^{2}_t - \frac{ \mu^{21}_t F^{2}_t}{2 T^{2}_t} dt \right)^2 }{4\left(\frac{d t}{4T^2_t} (\mu^{21}_t  )^2 +\frac{G^{11}_t}{2}\right)} \right].
\label{sup:probability}
\end{align}

From Eqs. (\ref{sup:Gaussian2}),  (\ref{sup:pathintegral}), and (\ref{sup:probability}), we obtain the analytical expression of the transfer entropy $d I^{\rm tr}_t$ up to the order of $dt$:
\begin{align}
d I^{\rm tr }_t := & \langle  \ln p [ x^{2}_{t+dt}| x^2_t, x^1_t] +\ln p [x^2_{t} ] - \ln p [x^2_{t+dt},  x^{2}_t ] \rangle \nonumber \\
=& -  \frac{dt}{4T^2_t} \langle \left( F^2_t -\mu^{21}_t \bar{x}^1_t  \right)^2 \rangle - \frac{1}{2}\ln \left[2\pi\Sigma^{22}_t \right]    - \frac{1}{2} (\Sigma^{22}_t)^{-1} \langle (\bar{x}^{2}_t)^2 \rangle + \frac{1}{2} \ln \left[4\pi \det{\Sigma_t }\left( \frac{d t}{4 T^2_t} (\mu^{2 1}_t )^2 +\frac{G^{11}_t}{2}\right) \right] \nonumber\\
& + \frac{d t}{4T^2_t} \langle (F^2_t)^2 \rangle + \frac{1}{2} G^{22}_t \langle (\bar{x}^{2}_t)^2  \rangle  - \frac{ \left< \left(  G^{12}_t \bar{x}^{2}_t - \frac{ \mu^{21}_t F^{2}_t}{2 T^{2}_t} dt \right)^2  \right>}{4\left(\frac{d t}{4T^2_t} (\mu^{21}_t  )^2 +\frac{G^{11}_t }{2}\right)}   \nonumber \\
=&\frac{\mu^{21} d t}{2T^2_t} \langle F^2_t \bar{x}^1_t\rangle - \frac{dt}{4T^2_t} ( \mu^{21}_t )^2 \Sigma^{11}_t -\frac{1}{2} +\frac{(\mu^{21}_t )^2  dt}{4 G^{11}_t T^2_t}    \nonumber  \\
&
+ \frac{1}{2} G^{22}_t \Sigma^{22}_t  -\frac{(G^{12}_t)^2 \Sigma^{22}_t}{2G^{11}_t}\left[ 1- \frac{d t}{2 G^{11}_t T^2_t} (\mu^{21}_t  )^2 \right]   +\frac{ \mu^{21}_t  dt}{2 G^{11}_t T_t^{2}} G^{12}_t  \langle F^{2}_t \bar{x}^{2}_t \rangle   - \frac{  (\mu^{21}_t )^2 d t}{4  G^{11}_t T^{2}_t }+\mathcal{O}(dt^2) \nonumber \\
=& \frac{\mu^{21} d t}{2T^2_t} \langle F^2_t \bar{x}^1_t\rangle +  \frac{ \mu^{21}_t  dt}{2 G^{11}_t T_t^{2}} G^{12}_t  \langle F^{2}_t \bar{x}^{2}_t \rangle   -  \frac{  (\mu^{21}_t )^2 d t}{4  G^{11}_t T^{2}_t }+\mathcal{O}(dt^2) \nonumber \\
=& \frac{(\mu^{21}_{t})^2 }{4T^2_t} \frac{\det{\Sigma_t}}{{\Sigma}^{22}_t} dt +\mathcal{O}(dt^2) \nonumber \\
=& \frac{1}{2} \ln \left( 1 + \frac{d{P}_{t}}{N_{t}} \right)  +\mathcal{O}(dt^2),
\label{analytical}
\end{align}
where we define $d{P}_t := (\mu^{21}_{t})^2 (\det{\Sigma_t }) dt/ (\Sigma^{22}_t)$, and  $N_t :=2T^{2}_t$. In this calculation, we used $ G^{ij}_t = G^{ji}_t$, $\Sigma^{ij}_t = \Sigma^{ji}_t$, $ G^{i1}_t \Sigma^{1l}_t +G^{i2}_t \Sigma^{2l}_t = \delta_{ij}$, $\langle (F^{2}_t)^2 \rangle dt^2 = 2 T^2_t dt +\mathcal{O}(dt^2) $, $\langle F^{2}_t \bar{x}^{1}_t\rangle = \mu^{21}_t \Sigma^{11}_t$,  $\langle F^{2}_t \bar{x}^{2}_t\rangle = \mu^{21}_t \Sigma^{12}_t$, and $G^{11}_t = (\Sigma_t^{22} )/(\det {\Sigma_t})$.

In the model of the {\it E. coli} bacterial chemotaxis, we have $N_t = 2T^m$ and
\begin{align}
dP_t &= \frac{1}{(\tau^m)^2} \frac{ [ \langle {a}_t^2 \rangle -\langle {a}_t \rangle^2 ][ \langle m_t^2 \rangle -\langle m_t \rangle^2 ] -[ \langle a_t m_t \rangle -\langle a_t \rangle \langle m_t \rangle ]^2   }{\langle m_t^2 \rangle -\langle m_t \rangle^2} dt\nonumber \\
&=  \frac{1- (\rho_t^{am})^2 }{(\tau^m)^2} V_t^a dt,
\label{sup:power}
\end{align} 
where $V^x_t := \langle x^2_t \rangle - \langle x_t \rangle^2$ indicates the variance of $x_t =a_t$ or $x_t =m_t$, and  $\rho_t^{am} :=  [\langle a_t m_t \rangle -\langle a_t \rangle \langle m_t \rangle] / (V^a_t V^m_t)^{1/2}$ is the correlation coefficient of $a_t$ and $m_t$. The correlation coefficient $\rho_t^{am}$ satisfies $-1 \leq \rho_t^{am} \leq 1$, because of the Cauchy-Schwartz inequality. We note that, if the joint probability $p(a_t,m_t)$ is Gaussian, the factor $1- (\rho_t^{am})^2$ can be rewritten by the mutual information $I^{am}_t$ as
\begin{align}
1- (\rho_t^{am})^2 = \exp [ -2 I^{am}_t],
\end{align} 
where $I^{am}_t$ is defined as $ I^{am}_t:= \int da_t dm_t p[a_t,m_t] \ln [p[ a_t,m_t] /[p[a_t]p[m_t]] ]$.
This fact implies that, if the target system $a_t$ and the other system $m_t$ are strongly correlated (i.e., $I^{am}_t \to \infty$), no information flow exists (i.e., $dI^{\rm tr}_t \to 0$).

From the analytical expression of the transfer entropy Eq.~(\ref{analytical}), we can analytically compare the conventional thermodynamic bound [i.e., $\Xi^{\rm SL}_t := - J^m_t dt /T^m_t+ dS^{am}_t \geq  J^a_t dt/ T^a_t$] with the information-thermodynamic bound (\ref{sup:infothermo}) for the model of {\it E. coli} chemotaxis [Eqs.~(\ref{sup:Langa}) and (\ref{sup:Langb}) with $ \bar{a}_t = \alpha m_t - \beta l_t$] in a stationary state, where both of the Shannon entropy and the conditional Shannon changes vanish, i.e., $dS^{a|m}_t=0$ and $dS^{am}_t=0$. Thus, the conventional thermodynamic bound is given by the heat emission from $m$ such that $\Xi^{\rm SL}_t = - J^m_t dt/T^m_t$, and the information thermodynamic bound is given by the information flow such that $\Xi^{\rm Info}_t = dI^{\rm tr}_t$. The information thermodynamic bound is given by $\Xi^{\rm Info}_t = (1- (\rho^{am}_t)^2) [\langle a_t^2\rangle -\langle a_t\rangle^2] dt /[2 (\tau^m)^2 T^m_t]$. The conventional thermodynamic bound is given by $\Xi^{\rm SL}_t =\langle a_t^2 \rangle dt/[ (\tau^m)^2 T^m_t]$. From $-1 \leq\rho^{am}_t \leq 1$ and $\langle a_t\rangle^2 \geq 0$, we have inequality $\Xi^{\rm SL}_t  \geq \Xi^{\rm Info}_t$. This implies that the information-thermodynamic bound $\Xi^{\rm Info}_t$ is tighter than the conventional bound $\Xi^{\rm SL}_t$ for the model of {\it E. coli} bacterial chemotaxis: 
\begin{align}
\Xi^{\rm SL}_t \geq \Xi^{\rm Info}_t \geq J^a_t dt /T^a_t.
\end{align}

\newpage
		 			\begin{figure*}[tb]
		\centering
		\includegraphics[width=180mm, bb=68 69 509 475]{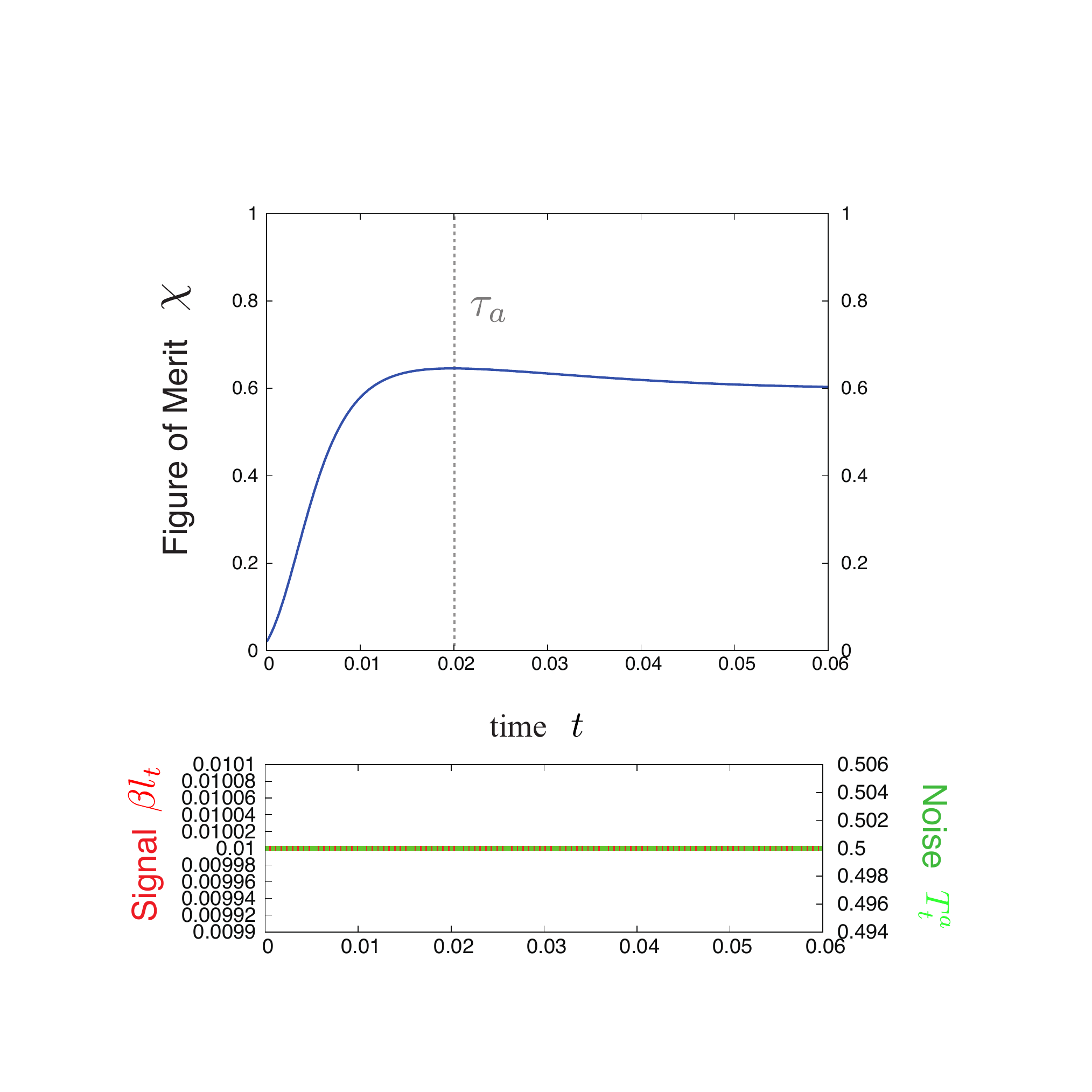}
		
		{\bf Supplementary Figure 1 $\mid$ A figure of merit of information thermodynamics: Step function.} The parameters are chosen as the same as in Fig. 2a in the main text.  
	\end{figure*}
	\newpage
	
		 			\begin{figure*}[tb]
		\centering
		\includegraphics[width=180mm, bb=86 87 497 498]{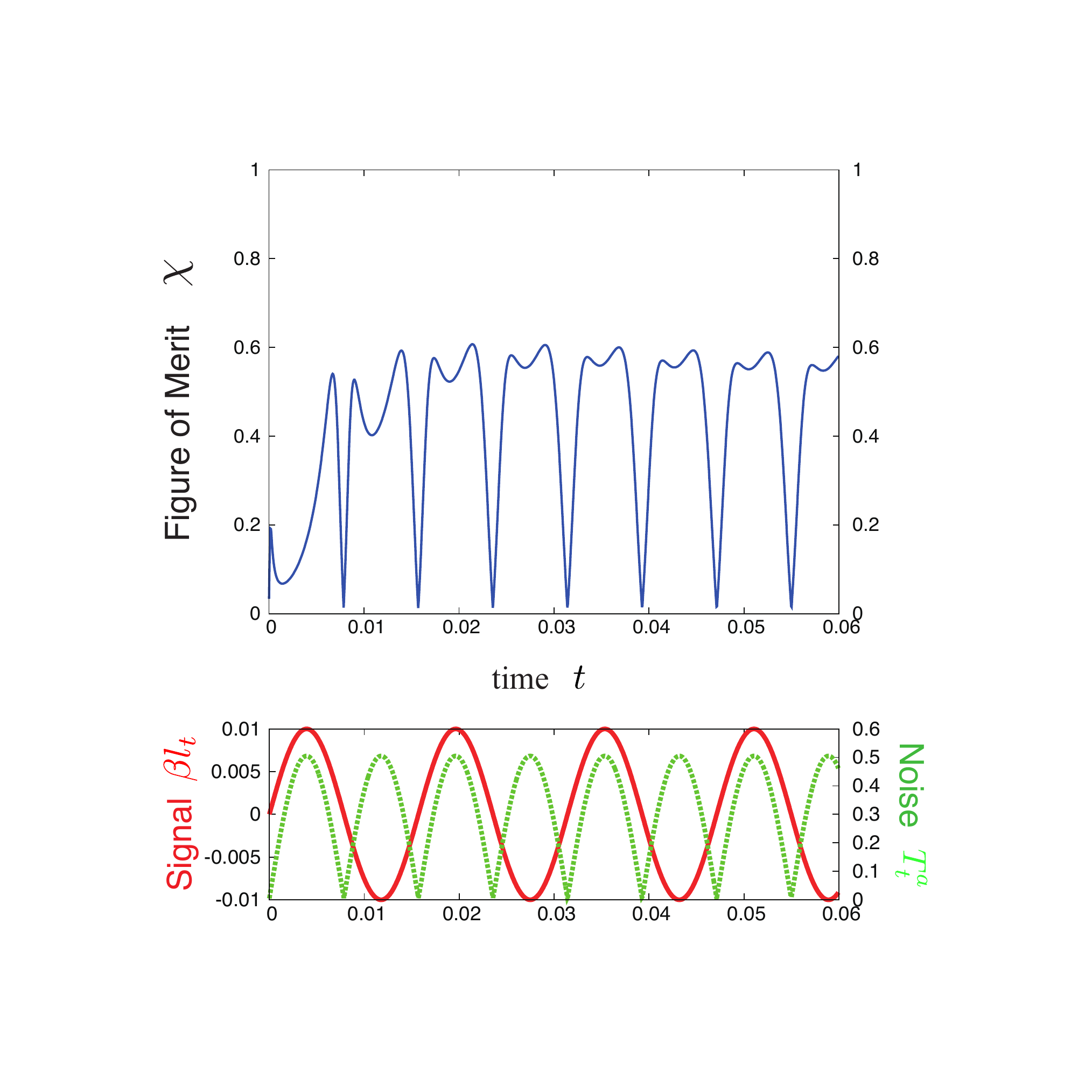}
		
		{\bf Supplementary Figure 2 $\mid$ A figure of merit of information thermodynamics: Sinusoidal function.} The parameters are chosen as the same as in Fig. 2b in the main text.
	\end{figure*}
	\newpage
	
		 			\begin{figure*}[tb]
		\centering
		\includegraphics[width=180mm, bb=74 140 485 553]{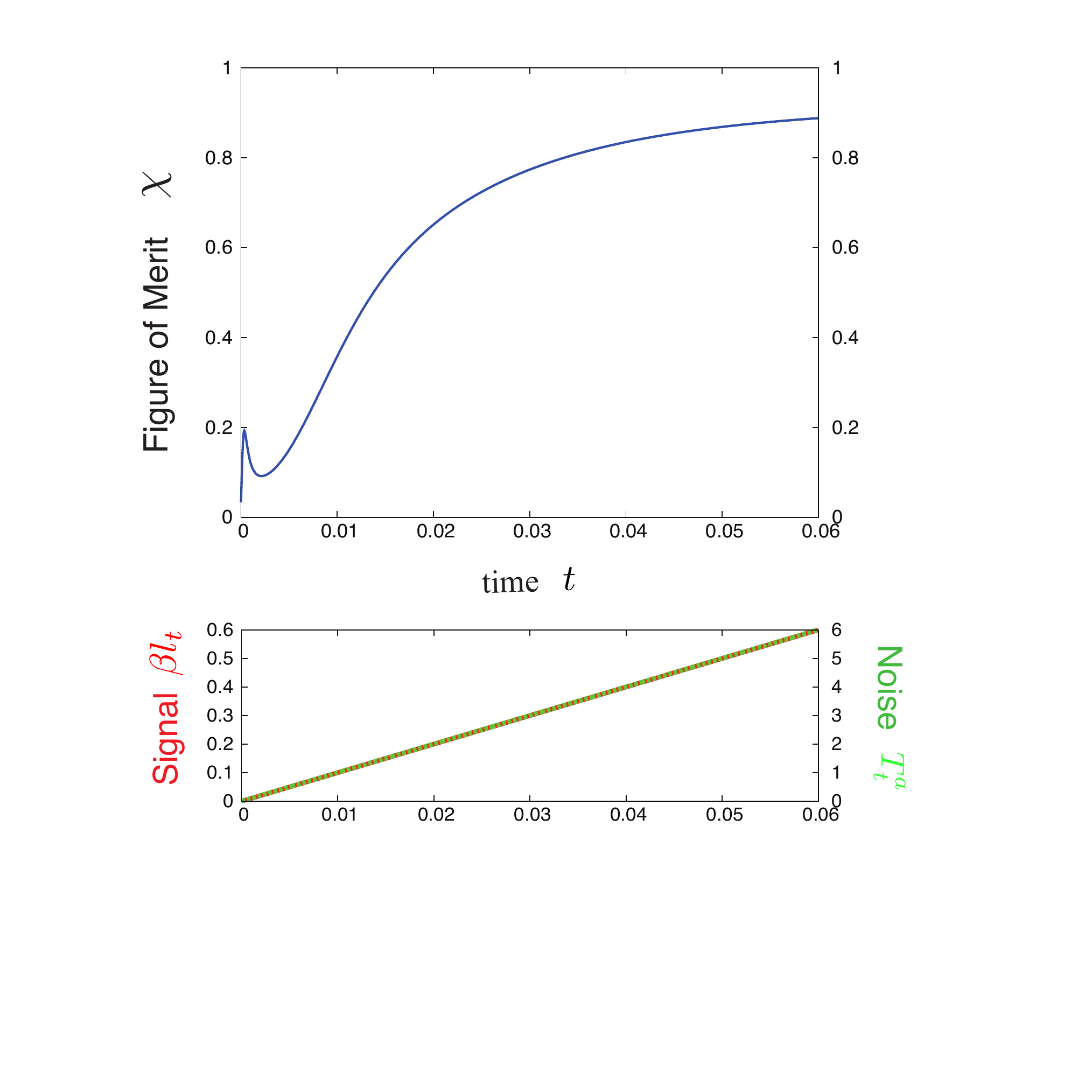}
		
		{\bf Supplementary Figure 3 $\mid$ A figure of merit of information thermodynamics: Linear function.}The parameters are chosen as the same as in Fig. 2c in the main text. 
	\end{figure*}
	\newpage
	
		\begin{figure*}[tb]
		\centering
		\includegraphics[width=180mm, bb=61 84 483 498]{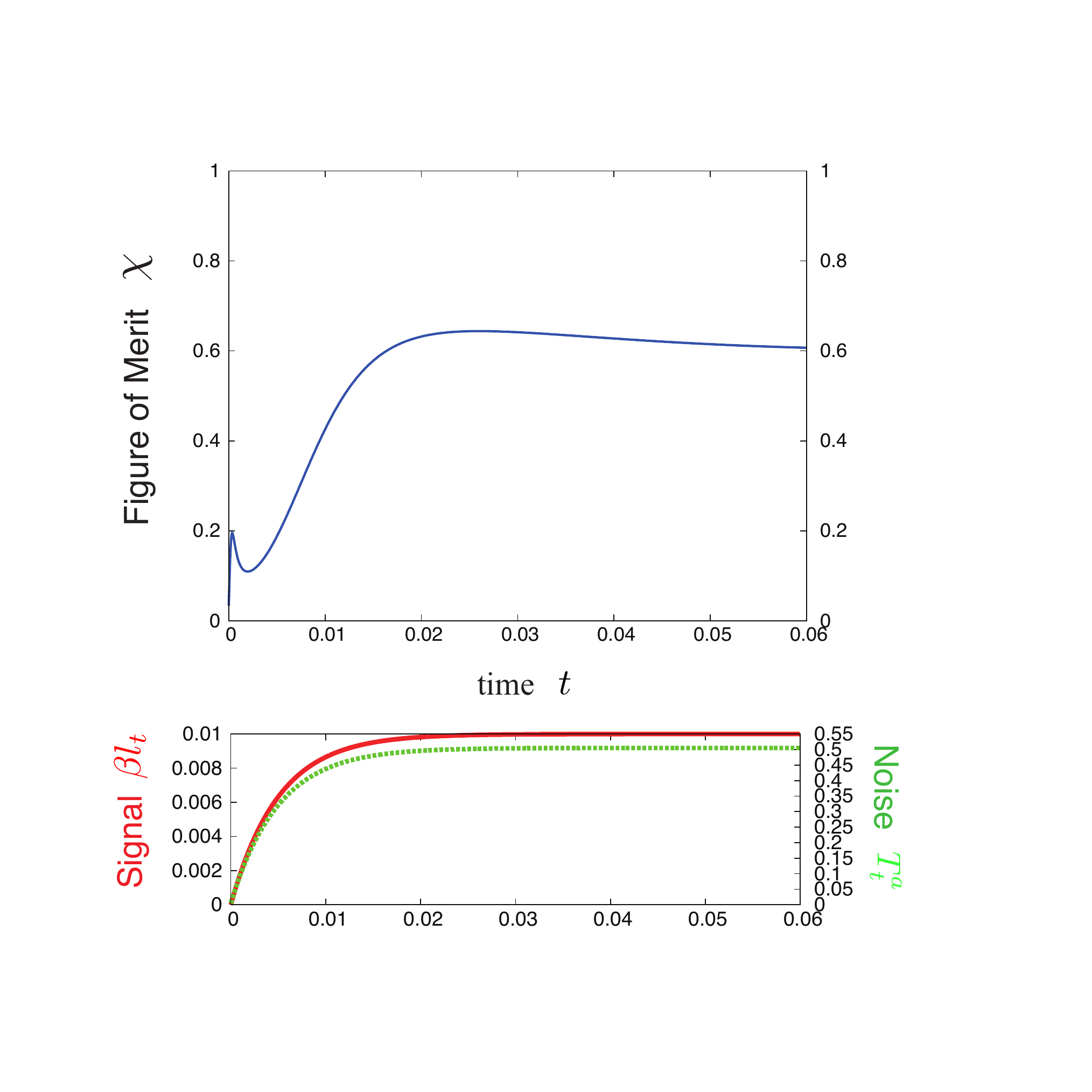}
		
		{\bf Supplementary Figure 4 $\mid$ A figure of merit of information thermodynamics: Exponential decay.} The parameters are chosen as the same as in Fig. 2d in the main text. 
	\end{figure*}
	
	\newpage
			\begin{figure*}[tb]
		\centering
		\includegraphics[width=180mm, bb=47 110 479 526]{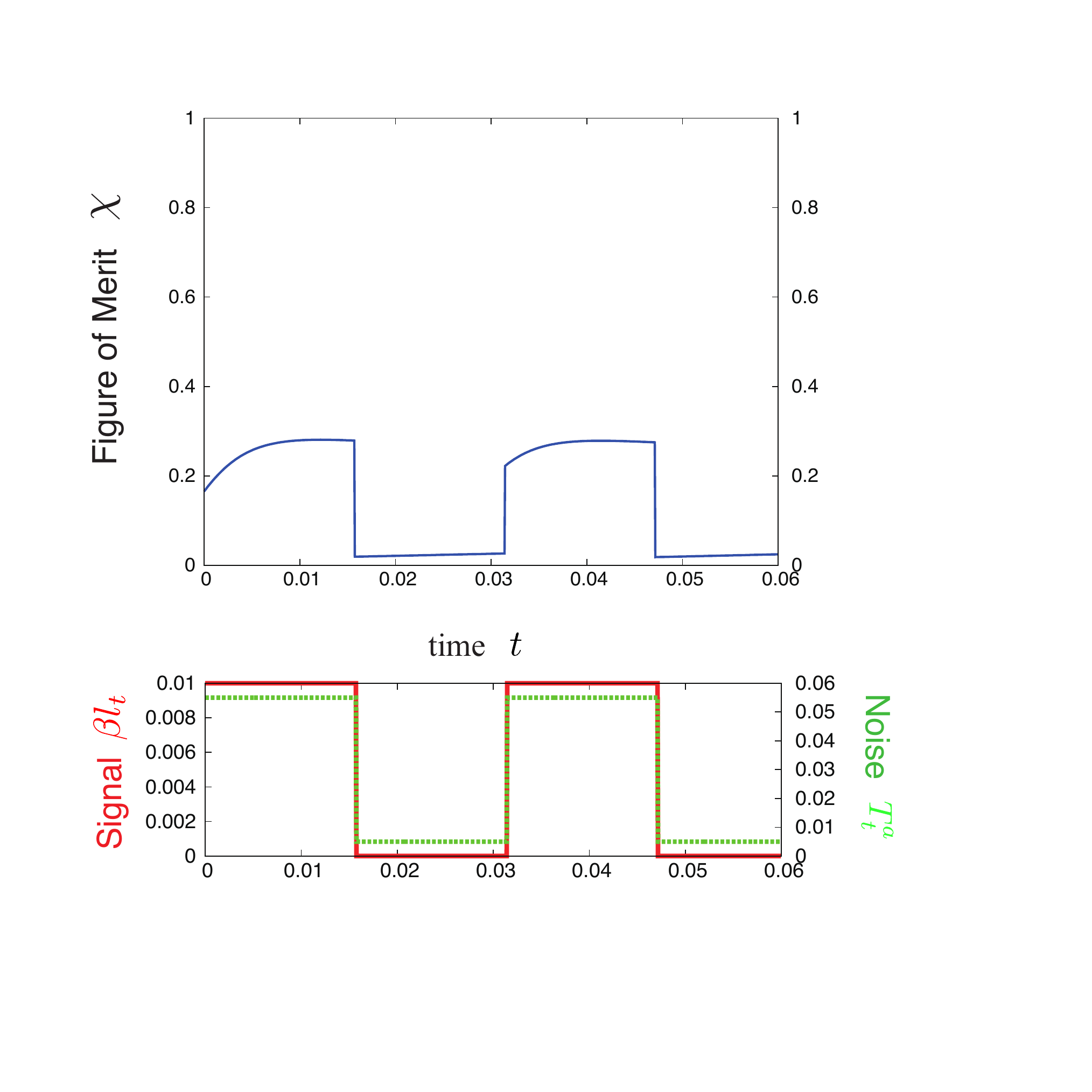}
		
		{\bf Supplementary Figure 5 $\mid$ A figure of merit of information thermodynamics: Square wave.} The parameters are chosen as the same as in Fig. 2e in the main text. 
	\end{figure*}
	
	\newpage
	\begin{figure*}[tb]
		\centering
		\includegraphics[width=180mm, bb=83 85 504 503]{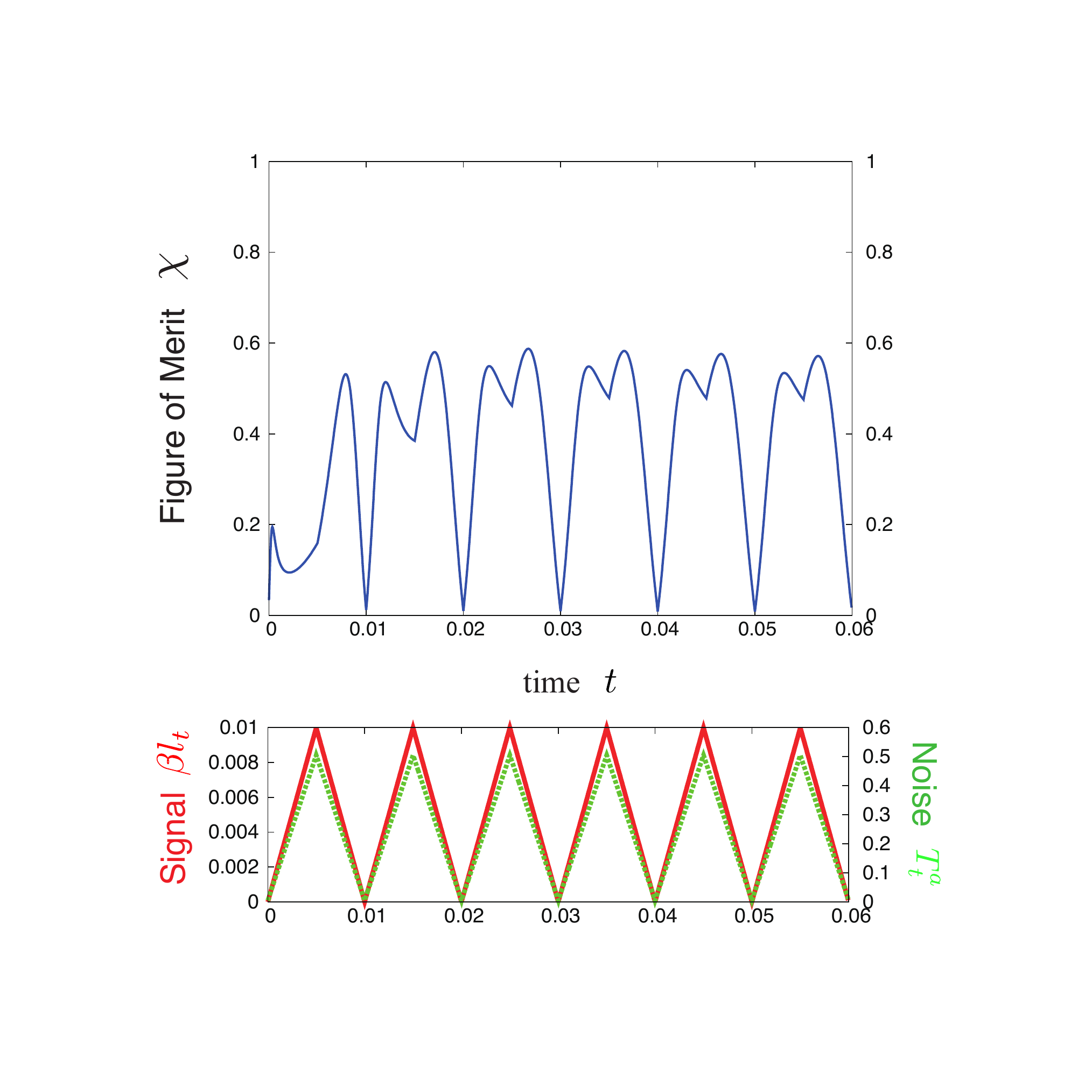}
		
		{\bf Supplementary Figure 6 $\mid$ A figure of merit of information thermodynamics: Triangle wave.} The parameters are chosen as the same as in Fig. 2f in the main text.  
	\end{figure*}
	\newpage
		\begin{figure*}[tb]
		\centering
		\includegraphics[width=180mm, bb=93 16 537 418]{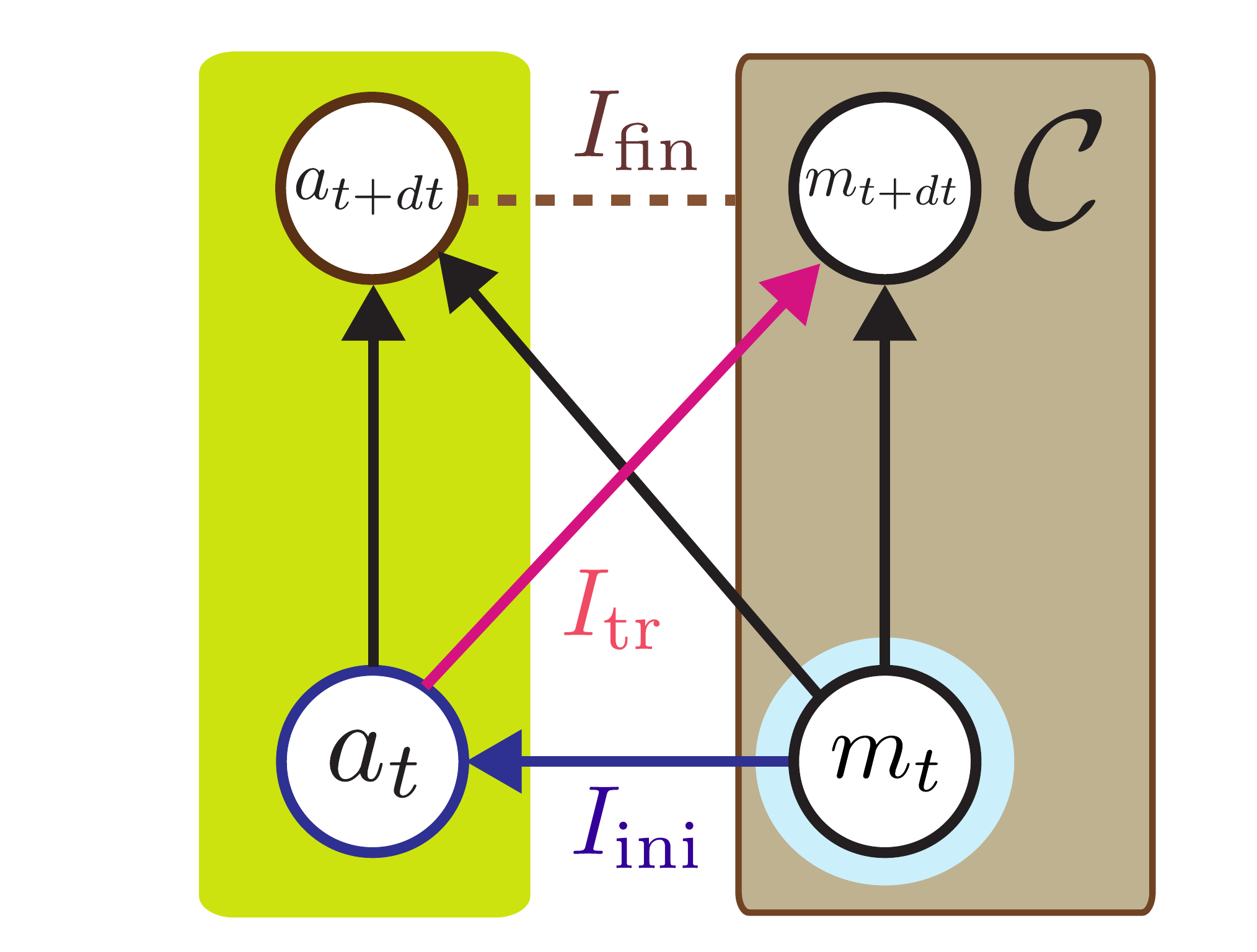}
		
		　
		
		{\bf Supplementary Figure 7 $\mid$ A Bayesian network corresponding to Eq. (\ref{sup:oneprocess}) in Supplementary note 3.} This Bayesian network gives the joint probability Eq. (2), where a node represents a random variable and an edge represent a causal relationship. Due to a general framework of information thermodynamics~\cite{ItoSagawa}, information of initial correlation $I_{\rm ini}$ is characterized by the mutual information between $a_t$ and $m_t$, the information of final correlation $I_{\rm fin}$  is characterized by the mutual information between $a_{t+dt}$ and $\{ m_t, m_{t+dt} \}$, and the transfer entropy $I_{\rm tr}$  from the subsystem a to the other system $\mathcal{C}$ is characterized by the conditional mutual information between $a_t$ and $m_{t+dt}$ under the condition of $m_t$. These information quantities $I_{\rm ini}$, $I_{\rm fin}$ , and $I_{\rm tr}$ give a lower bound of the entropy production in the subsystem $a$. 
	\end{figure*}

\end{document}